\documentclass[11pt,a4paper]{article}

\usepackage{jheppub} 

\usepackage{pstricks}
\usepackage{bigstrut}

\newcommand{\mre}{\mathrm{e}}

\newcommand{\mrd}{\mathrm{d}}
\newcommand{\N}{{\mathbb{N}}}
\newcommand{\Z}{{\mathbb{Z}}}
\newcommand{\R}{{\mathbb{R}}}

\newcommand{\phat}{\hat{p}}
\newcommand{\pz}{\partial_0}

\newcommand{\psump}{\sum_{p}\rule{0pt}{2.5ex}'\;}
\newcommand{\psum}[1]{\sum_{#1}\rule{0pt}{2.5ex}'\;}

\newcommand{\mcS}{\mathcal{S}}
\newcommand{\mcV}{\mathcal{V}}
\newcommand{\mcVD}{\mathcal{V}_D}
\newcommand{\mcVp}{\mathcal{V}'}

\newcommand{\mcHov}{\overline{\mathcal{H}}}
\newcommand{\mcW}{\mathcal{W}}
\newcommand{\mcWov}{\overline{\mathcal{W}}}
\newcommand{\mcZov}{\overline{\mathcal{Z}}}
\newcommand{\Cov}{\overline{C}}
\newcommand{\gov}{\overline{g}}

\newcommand{\Zov}{\overline{Z}}
\newcommand{\Wov}{\overline{W}}
\newcommand{\Psiov}{\overline{\Psi}}

\newcommand{\Ls}{L_s}

\newcommand{\Lhat}{\widehat{L}}
\newcommand{\ellhat}{\hat{\ell}}
\newcommand{\uell}{\underline{\ell}}
\newcommand{\IDR}{\overline{I}}
\newcommand{\bfp}{{\bf p}}
\newcommand{\bfx}{{\bf x}}
\newcommand{\bfy}{{\bf y}}
\newcommand{\bfz}{{\bf z}}

\newcommand{\phm}{\phantom{-}}
\newcommand{\phmz}{\phantom{-0}}
\newcommand{\phmzz}{\phantom{-00}}
\newcommand{\order}[1]{\mathrm{O}\left( #1 \right) }

\title{Massless sunset diagrams in finite asymmetric volumes}

\author[a]{F.\ Niedermayer}
\author[b]{and P.\ Weisz}

\affiliation[a]{Albert Einstein Center for Fundamental Physics, \\
Institute for Theoretical Physics, University of Bern, 
Switzerland}
\affiliation[b]{Max-Planck-Institut f\"ur Physik, 80805 Munich, Germany}

\emailAdd{niedermayer@itp.unibe.ch}
\emailAdd{pew@mpp.mpg.de}

\abstract{This paper discusses the methods and the results used
in an accompanying paper describing the matching of effective chiral
Lagrangians in dimensional and lattice regularizations.
We present methods to compute 2-loop massless sunset diagrams in finite 
asymmetric volumes  in the framework of these regularizations.
We also consider 1-loop sums in both regularizations,
extending the results of Hasenfratz and Leutwyler for the case of
dimensional regularization and we introduce a new method to calculate
precisely the expansion coefficients of the 1-loop 
lattice sums.}

\subheader{\hfill MPP-2015-267}

\begin{document} 

\maketitle

\section{Introduction}
\label{Introduction}

In an accompanying paper \cite{Nie15a} we have computed a free
energy and the mass gap in the $\delta$-regime
in the framework of chiral perturbation theory without
an explicit symmetry breaking term, in finite asymmetric volumes.
The computation was done both in dimensional regularization (DR)
and in the lattice regularization.
Matching the results in two regularizations enables us to 
establish relations between the 4-derivative couplings
appearing in these regularizations. 
The technical details are provided in the present paper.

In the computation 
of the free energy and the finite-volume mass gap
we encounter one and two loop integrals over the volume.  
A class of 1-loop sums with dimensional regularization
have been considered in detail by
Hasenfratz and Leutwyler \cite{Has90} in DR, we summarize some
of their results and also add some more.
The main part of the paper deals with the precise evaluation 
of the 2-loop sunset diagrams for both regularizations.
For our purposes we only require
diagrams with zero external momenta but the methods are
more general. Massive sunset diagrams in finite volume
have been treated in detail by Bijnens, Bostr\"{o}m and L\"{a}hde
\cite{Bij13,Bij14}, but as far as we can tell their methods are 
not applicable for zero masses.
In Appendix B we outline an alternative method of calculating
sunset diagrams by introducing a mass, in addition to the finite volume. 
This setup allows to take the massless limit as well.

Finally in sect.~6 we consider 1- and 2-loop massless
sums in asymmetric finite volumes with lattice
regularization. For the 1-loop sums we introduce a new
method to compute precisely the coefficients of
their expansions in the lattice cutoff.

The motivation to consider lattice-regularized chiral 
perturbation theory is addressed in \cite{Nie15a}.

\section{Dimensional regularization at finite volume}

Our goal is to compute massless one and two loop diagrams
in finite volume with periodic boundary conditions within
the framework of dimensional regularization. As in
\cite{Has90} we start by considering diagrams in 
the massive theory.

The free massive propagator is
\begin{equation}
G(x,M)=\frac{1}{V}\sum_p\frac{\mre^{ipx}}{p^2+M^2}\,,
\end{equation}
where $p$ runs over a $d$-dimensional momentum space infinite lattice
\begin{equation}
p=2\pi(n_0/L_0,\dots,n_{d_s}/L_{d_s})\,,
\end{equation}
where $d_s=d-1$ and $n_\mu$ are integers and
\begin{equation}
V=\prod_{\mu=0}^{d_s} L_\mu\,.
\end{equation}
The numerical evaluation of the graphs will be done for spatially
cubic volumes $L_1=\ldots=L_{d_s}\equiv L_s$, either for the hypercubic case
$L_0=L_s$ or for the elongated geometry $L_0 > L_s$.

We will dimensionally regularize by adding $q$ extra compact dimensions
of size $\Lhat$ and analytically continue the resulting loop formulae
to $q=-2\epsilon$.

\subsection{Massive propagator sums in DR }

In appendices of their classic paper \cite{Has90}
Hasenfratz and Leutwyler considered the sums
\begin{align}
  G_r&=\frac{1}{V_D}\sum_p H_r(p)\,,\label{Gr}
  \\
  H_r(p)&=\Gamma(r)(p^2+M^2)^{-r}\,,\label{Hr}
\end{align}
where $p$ is now a $D=d+q$ dimensional vector and
\begin{equation} 
  V_D\equiv V\Lhat^q\,.
\end{equation}
For convenience, in this subsection we reproduce some of their results,
in particular those that we need in \cite{Nie15a}.

Invoking the identity
\begin{equation}
  \frac{1}{V_D}\sum_p f(p)=\sum_w\widetilde{f}(w)
\end{equation}
where $\widetilde{f}(x)$ is the Fourier transform of $f(p)$:
\begin{equation}
  \widetilde{f}(x)=\int\frac{\mrd^Dp}{(2\pi)^D}\,\mre^{ipx}f(p)\,,
\end{equation}
and the sum over $w$ extends over the coordinate space lattice
\begin{equation}
  w=\left(v_0L_0,\dots,v_{D-1}L_{D-1}\right)\,,
\end{equation}
where $v_k$ are integers, one obtains
\begin{equation}
  G_r=\sum_w \widetilde{H}_r(w)\,,
\end{equation}
with
\begin{equation}
  \widetilde{H}_r(x)=\int_0^\infty\mrd\lambda\,\lambda^{r-1}(4\pi\lambda)^{-D/2}
  \exp\left(-\lambda M^2-\frac{x^2}{4\lambda}\right)\,.
\end{equation}
In the sum over $w$ the term $w=0$ corresponds to the infinite volume limit
\begin{equation}
  \widetilde{H}_r(0)=
  G_{r,\infty}=\Gamma(r)\int\frac{\mrd^D p}{(2\pi)^D}\,(p^2+M^2)^{-r}
  =\frac{\Gamma(r-D/2)}{(4\pi)^{D/2}}M^{D-2r}\,.
  \label{Grinfty}
\end{equation}
In particular 
\begin{equation}
  G_{1,\infty}=
  \begin{cases}
    -\frac{M^{D-2}}{2\pi}\left[\frac{1}{D-2}
      +\Cov+\frac12\right]+\order{D-2}\,\,\,\,\,&
    \text{for }D-2\sim0\,,\\
    -\frac{M}{4\pi}\,\,\,\,\,\,&\text{for }D=3\,,\\
    \frac{M^{D-2}}{8\pi^2}\left[\frac{1}{D-4}
      +\Cov\right]+\order{D-4}\,\,\,\,\,\,&\text{for }D-4\sim0\,,
  \end{cases}
\end{equation}
with
\begin{equation}
  \Cov=-\frac12\left[\ln(4\pi)-\gamma_E+1\right]\,,
\end{equation}
where $\gamma_E=-\Gamma'(1)=0.577\dots$ is the Euler constant.

It is useful to separate the infinite volume term to arrive at 
the representation
\begin{align}
  G_r&=G_{r,\infty}+g_r+\order{D-d}\,,
  \\
  g_r&=\int_0^\infty\mrd\lambda\,\lambda^{r-1}(4\pi\lambda)^{-d/2}
  \mre^{-\lambda M^2}
  \sum_{v\ne0}\exp\left(-\sum_{\mu=0}^{d_s}\frac{v_\mu^2L_\mu^2}{4\lambda}\right)\,.
\end{align}
In this decomposition the volume dependence is exclusively contained in
the function $g_r$ which is unambiguous for any integer value of $D$
and is entire in $r$. 

The dimensional regularization only affects the 
volume-independent part $G_{r,\infty}$ which contains poles at 
$D=2r,2r+2,\dots$.

The sum
occurring in the representation of $g_r$ converges rapidly if the sides 
of the box are large compared to the Compton wavelength, $ML_\mu\gg1$.
If all $L_\mu\to\infty$ then $g_r\to0$ exponentially.
Note if one considers other limits e.g.\ the temperature 
to $T=1/L_0$ to zero i.e. $L_0\to\infty$ with $L_1,\dots,L_{d_s}$ fixed,
then $g_r$ approaches a finite limit
\begin{equation}
  \lim_{L_0\to\infty}g_r^{(d)}=(4\pi)^{-1/2}g_{r-1/2}^{(d-1)}\,.
\end{equation}
For $r=0$ this relation reflects the fact that in the limit $T\to0$
the partition function is dominated by the contribution from the
ground state, the quantity $\ln\det D/2L_0$ approaching the Casimir
energy associated with the $d_s-$dimensional box.

For the goal of considering the massless sums we need
the properties of $g_r$ in the limit $M\to0$ where
infrared singularities occur. 
Introducing 
\begin{equation} \label{Sx}
  S(x)\equiv\sum_{n=-\infty}^\infty\mre^{-\pi n^2 x}\,,\,\,\,\,\,\,(x>0)\,,
\end{equation}
(related to the Jacobi theta-function, defined in \eqref{Suz}),
we have  
\begin{equation}
  g_r=\frac{1}{L^d}\left(\frac{L^2}{4\pi}\right)^r\int_0^\infty\mrd t\, 
  t^{r-d/2-1}
  \exp\left(-\frac{z^2t}{4\pi}\right)
  \left[\prod_{\mu=0}^{d_s}\,S\left(\frac{\ell_\mu^2}{t}\right)-1\right]\,,
  \label{gr}
\end{equation}
where $L$ is some reference scale,
\begin{equation}
  \ell_\mu\equiv L_\mu/L\,,
\end{equation}
and
\begin{equation}
  z\equiv ML\,.
\end{equation}
We will use the shorthand notation 
$\uell = \{\ell_0,\ell_1,\ldots,\ell_{d_s} \}$ for the relative sizes
in the physical dimensions and
$\ellhat=\Lhat/L=L_\alpha/L$ for $\alpha=d,\ldots,D-1$
for the extra dimensions.

The dependence on the auxiliary scale $L$ of a given quantity 
(when the other variables are made dimensionless) is given by its dimension.
Using this we will often use $L=L_s$ for the spatially cubic volume
and denote $\ell\equiv\ell_0=L_0/L_s$ and $\ellhat =\Lhat/L_s$.

Since the product in \eqref{gr} often appears below we also
introduce the ancillary definition (with some abuse of notation)
\begin{equation}
  \mcS_d\left( \frac{\ell^2}{t} \right) 
  \equiv \prod_{\mu=0}^{d-1}\,S\left(\frac{\ell_\mu^2}{t}\right)\,,
  \qquad
  \mcS_d\left( \frac{t}{\ell^2} \right) 
  \equiv \prod_{\mu=0}^{d-1}\,S\left(\frac{t}{\ell_\mu^2}\right)\,.
  \label{Sddef}
\end{equation}
(For the general case of $D$ dimensions we shall also use
$\mcS_D(t/\ell^2)$ and $\mcS_D(\ell^2/t)$ defined analogously.)

Note that the sum in \eqref{Sx} converges very fast for $x>1$.
The function $S(x)$ satisfies the well known remarkable identity
\begin{equation}
  S(x)=x^{-1/2}S(x^{-1})\,.  
  \label{Sxi}
\end{equation}
Using \eqref{Sx} for $x>1$ and \eqref{Sxi} for $x<1$ one needs
only a few terms in the corresponding sums to calculate $S(x)$
very precisely. 

Using \eqref{Sxi} one obtains the representation
\begin{equation}
  g_r = \frac{1}{L^d}\left(\frac{L^2}{4\pi}\right)^r
  \left[A_r+\mcV^{-1}B_r-B_{r-d/2}\right]\,,
\end{equation}
where 
\begin{equation}
  A_r = \frac{1}{\mcV}\int_0^\infty \mrd t \, t^{r-1}
  \exp\left(-\frac{z^2 t}{4\pi}\right)
  \left[\mcS_d\left(\frac{t}{\ell^2}\right)
  \right]_\mathrm{sub} \,,
  \label{ArA}
\end{equation}
\begin{equation}
  B_r = \int_1^\infty\mrd t\, t^{r-1} \exp\left(-\frac{z^2t}{4\pi}\right)\,, 
\end{equation}

and
\begin{equation}
  \mcV\equiv\prod_{\mu=0}^{d_s}\ell_\mu = V L^{-d} \,.
\end{equation}

In \eqref{ArA} we introduced the notation $[\ldots]_\mathrm{sub}$
which will be used often below. It is defined by 
\begin{equation}
  \left[ \Phi(t;\ldots) \right]_\mathrm{sub}
  = \Phi(t;\ldots) -
  \begin{cases}
    [\Phi(t;\ldots)]_0\,, & \text{for } 0<t<1\,, \\
    [\Phi(t;\ldots)]_\infty\,, & \text{for } 1<t<\infty\,, \\
  \end{cases}
  \label{Phisub}
\end{equation}
where $\left[\Phi(t;\ldots)\right]_0$ and 
$\left[\Phi(t;\ldots)\right]_\infty$ denote
the leading asymptotic parts of $\Phi(t;\ldots)$ 
at $t\to 0$ and $t\to\infty$, respectively.
According to \eqref{Sx} and \eqref{Sxi} one has
$\left[S(x)\right]_0=x^{-1/2}$ and $\left[S(x)\right]_\infty=1$.
In particular, in \eqref{ArA} one has
\begin{equation}
 \left[ \mcS_d\left(\frac{t}{\ell^2}\right)
  \right]_0 = \mcV\, t^{-d/2} \,,
  \qquad 
 \left[ \mcS_d\left(\frac{t}{\ell^2}\right)
  \right]_\infty = 1 \,.
\end{equation}

The quantity $A_r$ does not contain infrared singularities 
and has an expansion in $M^2$ of the form 
\begin{equation}
A_r=\sum_{n=0}^\infty\left(-\frac{z^2}{4\pi}\right)^n
\frac{1}{n!}\alpha_{r+n}\,.
\end{equation}
The expansion coefficients\footnote{This expression 
is equivalent to the one given in \cite{Has90}
but has a more compact form.} depend on $d$ and the ratios $\ell_\mu$:
\begin{equation}
  \alpha_s = \frac{1}{\mcV} \int_0^\infty \mrd t\, t^{s-1}
  \left[ \mcS_d\left(\frac{t}{\ell^2}\right)
  \right]_{\mathrm{sub}}  \,.
  \label{alphadef}
\end{equation}

Some values for $\alpha_s$ for $\ell_\mu=1$, $\forall \mu$ are 
given in table~\ref{alpha_leq1}.

\begin{table}[ht]
\centering
\begin{tabular}[t]{|l|c|c|c|c|}
\hline \bigstrut[t]
$d$ &$1$&$2$&$3$&$4$\\[1.0ex]
\hline 
$\alpha_{-1/2}$ &$0.04719755120$&$0.1044122116$&$0.1750738214$&$0.2641535689$\\
$\alpha_0$     &$0.04619141793$&$0.1008796989$&$0.1667047726$&$0.2474072414$\\
$\alpha_{1/2}$  &$0.04619141793$&$0.0997350800$&$0.1627025205$&$0.2379659789$\\
$\alpha_1$     &$0.04719755120$&$0.1008796989$&$0.1627025205$&$0.2349151988$\\
$\alpha_{3/2}$  &$0.04929326270$&$0.1044122116$&$0.1667047726$&$0.2379659789$\\
$\alpha_2$     &$0.05265787558$&$0.1106437295$&$0.1750738214$&$0.2474072414$\\
\hline
\end{tabular}
\caption{\footnotesize Values for $\alpha_s$ for $\ell_\mu=1\,\,\forall \mu$.
} 
\label{alpha_leq1}
\end{table}

Infrared divergences are contained in the incomplete 
$\Gamma$--function $B_s$, in the form of fractional powers of $M^2$:
\begin{equation}
B_s=\left(\frac{z^2}{4\pi}\right)^{-s}\Gamma(s)
-\sum_{n=0}^\infty\frac{1}{n!}
\left(-\frac{z^2}{4\pi}\right)^{n}\frac{1}{n+s}\,.
\end{equation}
The pole in the $\Gamma$--function at $r=-N\,,\,\,N\in\N$
is canceled by a pole occurring in the piece which is analytic in $M$.
Merging the two singularities one obtains a logarithmic contribution
\begin{equation}
  \begin{aligned}
    B_{-N}&=\frac{(-1)^{N+1}}{N!}\left(\frac{z^2}{4\pi}\right)^N
    \left\{\ln\left(\frac{z^2}{4\pi}\right)+\gamma_E
      -\sum_{n=1}^N\frac{1}{n}\right\}
    \\
    & \quad +\sum_{n=0,n\ne N}^\infty\frac{1}{n!}
    \left(-\frac{z^2}{4\pi}\right)^{n}\frac{1}{N-n}\,,\,\,\,\,N=0,1,2,\dots
  \end{aligned}
  \label{Bs}
\end{equation}
For $z=0$ and $s<0$ one has
\begin{equation}
  \left. B_{s}\right|_{z=0} = -\frac{1}{s}\,, \qquad \text{for } s<0\,.
  \label{Bz0}
\end{equation}

The small $M$ expansions of $g_r$ for integer $r$ can be obtained from
that of $g_0$ using the recursion relations
\begin{equation}
  g_{r+1}=-\mrd g_r/\mrd M^2\,.
  \label{grplus1}
\end{equation}

For $d=2$ the expansion of $g_0$ takes the form
\begin{align} \label{g0d2}
  g_0&=\frac{1}{L^2}\left[-2\mcV^{-1}\ln(z)
    -\frac{z^2}{2\pi}\ln(z)
    +\sum_{n=0}^\infty\frac{1}{n!}\beta_n z^{2n}\right]\,,
  \\
  \beta_0&=\alpha_0-1+\mcV^{-1}\left\{\ln(4\pi)-\gamma_E\right\}\,,
  \\
  \beta_1&=\frac{1}{4\pi}\left[-\alpha_1
    -\left\{-\ln(4\pi)+\gamma_E-1\right\}+\mcV^{-1}\right]\,,
  \\
  \beta_n&=\left(-\frac{1}{4\pi}\right)^n\left[
    \alpha_n+\frac{1}{n-1}-\frac{1}{n\mcV}\right]\,,\,\,\,n\ge2\,,d=2\,.
\end{align}

For $d=3$:
\begin{align}
  g_0&=\frac{1}{L^3}\left[-2\mcV^{-1}\ln(z)-\frac{z^3}{6\pi}
    +\sum_{n=0}^\infty\frac{1}{n!}\beta_n z^{2n}\right]\,, \label{g0d3}
  \\
  \beta_0&=\alpha_0-\frac23+\mcV^{-1}\left\{\ln(4\pi)-\gamma_E\right\}\,,
  \label{beta0d3}
  \\ 
  \beta_n&=\left(-\frac{1}{4\pi}\right)^n\left[\alpha_n+\frac{2}{2n-3}
    -\frac{1}{n\mcV}\right]\,,
  \,\,\,\,n\ne0\,,\,\,\,\,d=3\,.\label{betand3}
\end{align}

For $d=4$:
\begin{align}
  g_0&=\frac{1}{L^4}\left[-2\mcV^{-1}\ln(z)
    +\frac{z^4}{16\pi^2}\left\{\ln(z)-\frac14\right\}
    +\sum_{n=0}^\infty\frac{1}{n!}\beta_n z^{2n}\right]\,, \label{g0d4}
  \\
  \beta_0&=\alpha_0-\frac12+\mcV^{-1}\left\{\ln(4\pi)-\gamma_E\right\}\,,
  \\
  \beta_2&=\frac{1}{16\pi^2}\left[\alpha_2-\ln(4\pi)+\gamma_E-1
    -\frac{1}{2\mcV}\right]\,,
  \\
  \beta_n&=\left(-\frac{1}{4\pi}\right)^n\left[\alpha_n+\frac{1}{n-2}
    -\frac{1}{n\mcV}\right]\,,\,\,\,\,n\ne0,2\,,\,\,\,\,\,\,\,d=4\,.
  \label{betand4}
\end{align}

The table~\ref{beta_leq1} gives results for the symmetric box
$\ell_\mu=1\,\,\forall \mu$ and table~\ref{beta_leq2} for
$\ell=2$ for $d=3,4$. 
\begin{table}[ht]
\centering
\begin{tabular}[t]{|l|c|c|}
\hline  \bigstrut[t]
$d$ &$3$&$4$\\[1.0ex]
\hline 
$\beta_0$&$1.45384668796818338855$&$\phm 1.70121582349712182477$\\
$\beta_1$&$0.22578495944075803348$&$\phm 0.14046098554536575282$\\
$\beta_2$&$0.01060752889198424526$&$-0.02030477369161629693$\\
\hline
\end{tabular}
\caption{\footnotesize Values for $\beta_r$ for $\ell_\mu=1\,\,\forall \mu$
for $d=3,4$.} 
\label{beta_leq1}
\end{table}

\begin{table}[ht]
\centering
\begin{tabular}[t]{|l|c|c|}
\hline \bigstrut[t]
$d$ &$3$&$4$\\[1.0ex]
\hline 
$\beta_0$&$0.74461239033155890201$&$\phm 0.98194779750230477518$\\
$\beta_1$&$0.14370432528775141208$&$\phm 0.05911493648278131899$\\
$\beta_2$&$0.02021612362190113525$&$-0.01075957063969698115$\\
\hline                              
\end{tabular}
\caption{\footnotesize Values for $\beta_r$ for $\ell_\mu=1\,,\,\,\mu>0$
and $\ell=2$\,.} 
\label{beta_leq2}
\end{table}

\subsubsection{Sums with extra factor in the numerator}

Next we apply the methods of Hasenfratz and Leutwyler to the sums 
\begin{equation}
  G_{r,1}=\frac{1}{V_D}\sum_p p_\nu^2 H_r(p)\,,
  \,\,\,{\rm no\,\,sum\,\,over}\,\,\nu\,.
  \label{Grc1}
\end{equation}
We have
\begin{equation}
  G_{r,1}=G_{r,1,\infty}+g_{r,1}+\order{D-d}\,,
\end{equation}
where for infinite volume:
\begin{equation}
  G_{r,1,\infty}=\Gamma(r)\int\frac{\mrd^D p}{(2\pi)^D}\,p_\nu^2(p^2+M^2)^{-r}
  =\frac{\Gamma(r-1-D/2)}{2(4\pi)^{D/2}}M^{D-2r+2}=\frac12 G_{r-1,\infty}\,,
\end{equation}
and
\begin{align}
  g_{r,1}&=\frac12 g_{r-1} + \ell_\nu^2 h_r\,, \label{grc1A}
  \\
  h_r&= -\frac14 L^2
  \int_0^\infty\mrd\lambda\,\lambda^{r-3}(4\pi\lambda)^{-d/2}\mre^{-\lambda M^2}
  \sum_v v_\nu^2\exp\left(-\sum_{\mu=0}^{d_s}
    \frac{v_\mu^2L_\mu^2}{4\lambda}\right)\,. \label{hrA}
\end{align}
Denote the logarithmic derivative of $S(x)$ by
\begin{equation}
  T(x) = \frac{S'(x)}{S(x)} = -\frac{\pi}{S(x)} 
  \sum_{n=-\infty}^\infty n^2 \mre^{-\pi n^2 x} \,,
  \label{Tdef}
\end{equation}
satisfying the relation
\begin{equation}
  T(x) = -\frac{1}{2x} - \frac{1}{x^2}T\left(\frac{1}{x}\right) \,.
  \label{Trel}
\end{equation}
With this one has
\begin{equation}
  h_r=\frac{1}{L^d}\left(\frac{L^2}{4\pi}\right)^{r-1}
  \int_0^\infty\mrd t\,t^{r-d/2-3}\exp\left(-\frac{z^2t}{4\pi}\right)
  T\left(\frac{\ell_\nu^2}{t}\right)
  \mcS_d\left(\frac{\ell^2}{t}\right)\,.
\end{equation}

Putting the parts together we have
\begin{equation}
  g_{r,1}=\frac12\frac{1}{L^d}\left(\frac{L^2}{4\pi}\right)^{r-1}
  \left[-B_{r-1-d/2}+D_r\right]\,,
  \label{gr1}
\end{equation}
where $D_r$ has the expansion in $z$:
\begin{equation}
  D_r=\sum_{n=0}^\infty\left(-\frac{z^2}{4\pi}\right)^n
  \frac{1}{n!}\gamma_{r+n}\,,
  \label{Dr}
\end{equation}
with
\begin{equation}
  \gamma_s = - \frac{2}{\ell_\nu^2 \mcV}\int_0^\infty
  \mrd t\, t^{s-1}\left[
    T\left( \frac{t}{\ell_\nu^2} \right)
    \mcS_d\left( \frac{t}{\ell^2} \right)
  \right]_{\mathrm{sub}} \,.
  \label{gammadef}
\end{equation}
For the definition of $[\ldots]_{\mathrm{sub}}$ see \eqref{Phisub}.
Note that $[T(x)]_0=-1/(2x)$ and $[T(x)]_\infty=0$.
 
For the totally symmetric box $\ell_\mu=1\,, \,\,\forall\,\mu$ we have
\begin{equation}
  \gamma_s = \frac{2}{d}(s-1) \alpha_{s-1}\,, \qquad \ell=1 \,.
  \label{gammas_leq1}
\end{equation}

We then have for $D\sim3$:
\begin{equation}
  G_{2,1}=\frac{1}{8\pi L}\left[\gamma_2-2\right]+\order{D-3}+\order{M}\,,
  \label{G21}
\end{equation}
and for $D\sim4$:
\begin{equation}
  \begin{aligned}
    G_{2,1}&=\frac{1}{8\pi L^2}\left[\gamma_2-1\right]
    \\
    &+\frac{M^2}{16\pi^2}\left[-\ln L+\frac{1}{D-4}-\frac12\gamma_3\right]
    +\order{D-4}+\order{M^4}\,,
  \end{aligned}
\end{equation}
(note $\gamma_2$ depends on $d$).

\subsubsection{Keeping first order $D-d$ terms in $G_{r,1}$}

In our computation \cite{Nie15a} we need to keep in consideration 
order $q=D-d$ terms in $G_{r,1}$:
\begin{equation}
  G_{r,1}=G_{r,1,\infty}+\mathcal{G}_{r,1}\,,
\end{equation}
where
\begin{equation}
  \mathcal{G}_{r,1}=\frac{1}{L^D}\left(\frac{L^2}{4\pi}\right)^{r-1}
  \left[\frac12\mathcal{G}_r+\ell_\nu^2\mcHov_r\right]\,,
\end{equation}
with
\begin{align}
  \mathcal{G}_r & =\int_0^\infty\mrd t\, t^{r-D/2-2}
  \exp\left(-\frac{z^2t}{4\pi}\right)
  \left[S\left(\frac{\ellhat^2}{t}\right)^q
    \mcS_d\left(\frac{\ell^2}{t}\right)-1\right]\,,
  \\
  \mcHov_r & =  \int_0^\infty\mrd t\, t^{r-D/2-3}
  \exp\left(-\frac{z^2t}{4\pi}\right)
  S\left(\frac{\ellhat^2}{t}\right)^q
  T\left(\frac{\ell_\nu^2}{t}\right)
  \mcS_d\left(\frac{\ell^2}{t}\right)\,.
\end{align}

So
\begin{equation}
  \mathcal{G}_{r,1}=(1-q\ln L)g_{r,1}(\uell)  +
  q\frac{1}{L^d}\left(\frac{L^2}{4\pi}\right)^{r-1}g_{r,1}^{(1)}
  (\uell,\ellhat) +\order{q^2}\,,
\end{equation}
with
\begin{equation}
  \begin{aligned}
    g_{r,1}^{(1)}&=-\frac{1}{\ell_\nu^2 \mcV}\int_0^\infty\mrd t\, t^{r-1}
    \exp\left(-\frac{z^2t}{4\pi}\right)
    \left[ \ln\left(\frac{1}{\ellhat}S\left(\frac{t}{\ellhat^2}\right)\right)
      T\left(\frac{t}{\ell_\nu^2}\right)
      \mcS_d\left(\frac{t}{\ell^2}\right)
    \right]_\mathrm{sub}
    \\
    & \quad + \frac14 \int_1^\infty\mrd t\, t^{r-d/2-2} \ln(t)
    \exp\left(-\frac{z^2t}{4\pi}\right) \,.
  \end{aligned}
  \label{gr11A}
\end{equation}
For the expression appearing here one has
$[\ldots]_0=\frac14 \ell_\nu^2 \mcV \, t^{-d/2-1}\ln(t)$
and $[\ldots]_\infty=0$.

We introduce
\begin{align}
  \rho_r(\uell,\ellhat) & = 
  \frac{1}{\mcV} \int_0^\infty \mrd t \, t^{r-1}
  \left[ \ln\left(\frac{1}{\ellhat} S\left(\frac{t}{\ellhat^2}\right) \right)
    \mcS_d\left(\frac{t}{\ell^2}\right) 
  \right]_\mathrm{sub} \,,
  \label{rhodef}
  \\
  \tau_r(\uell,\ellhat) & = 
  -\frac{1}{\ell_\nu^2 \mcV}\int_0^\infty\mrd t\, t^{r-1}
  \left[ \ln\left(\frac{1}{\ellhat}S\left(\frac{t}{\ellhat^2}\right)\right)
    T\left(\frac{t}{\ell_\nu^2}\right)
    \mcS_d\left(\frac{t}{\ell^2}\right) 
  \right]_\mathrm{sub} \,.
  \label{taudef}
\end{align}
They have a finite limit for $\ellhat\to\infty$
\begin{align}
  \rho_r(\uell,\infty) & = 
  -\frac{1}{2 \mcV} \int_0^\infty \mrd t \, t^{r-1} \ln(t) 
  \left[ 
    \mcS_d\left(\frac{t}{\ell^2}\right) 
  \right]_\mathrm{sub} \,,
  \label{rhoinf}
  \\
  \tau_r(\uell,\infty) & = 
  \frac{1}{2\ell_\nu^2 \mcV}\int_0^\infty\mrd t\, t^{r-1} \ln(t) 
  \left[ T\left(\frac{t}{\ell_\nu^2}\right)
    \mcS_d\left(\frac{t}{\ell^2}\right) 
  \right]_\mathrm{sub} \,.
  \label{tauinf}
\end{align}

Using these for $z=0$ and $r<d/2+1$ one obtains
\begin{equation}
  \left. g_{r,1}^{(1)} \right|_{z=0} = \tau_r(\uell,\ellhat)
  + \frac{1}{(d+2-2r)^2} \,.
  \label{gr11_z0}
\end{equation}

In particular, for the case $r=1$ (cf. \eqref{gr1}, \eqref{Dr})
\begin{equation}
  g_{1,1}|_{z=0}=\frac{1}{2L^d}\left[\gamma_1-\frac{2}{d}\right]\,,
  \label{g11z0}
\end{equation}
and for $r=1,d=4$:\footnote{The notation $\mcW_1$ was used in \cite{Nie15a}.}
\begin{equation}
  \left. g_{1,1}^{(1)} \right|_{z=0,d=4} \equiv \mcW_1(\uell,\ellhat)
  = \left. \tau_1(\uell,\ellhat) \right|_{d=4} + \frac{1}{16} \,.
  \label{calW1A}
\end{equation}
For $\ellhat\to\infty$ one obtains
\begin{equation}
  \mcW_1(\uell,\infty)
  = \left. \tau_1(\uell,\infty) \right|_{d=4} + \frac{1}{16} \,.
  \label{calW1infA}
\end{equation}

\section{Massless propagator sums 
in dimensional regularization}
\label{psumDR}

We are now in a position to give results for dimensionally
regularized momentum sums 
\begin{equation}
  \IDR_{nm}\equiv\frac{1}{V_D}\psump\frac{p_0^{2m}}{(p^2)^n}\,,
\end{equation}
which can be re-expressed in terms of the massless propagator 
\begin{equation}
  G(x)=\frac{1}{V_D}\psump\frac{\mre^{ipx}}{p^2}\,,
  \label{Gmass0}
\end{equation}
where the sum is over momenta 
$p=2\pi(n_0/L_0,\dots,n_{D-1}/L_{D-1})\,,\,\,n_k\in\Z\,,$ and the prime
on the sum means that the zero momentum is omitted: 
$\sum'_p=\sum_{p\ne0}$. 

We have e.g.\footnote{One has 
$-\Box G(x)=\delta^{(D)}(x)-1/V_D$
hence $\IDR_{00}=-\Box G(0)=-1/V_D$ where we used that $\delta^{(D)}(0)=0$
in DR.}
\begin{align}
  \IDR_{00}&=-\frac{1}{V_D}\,,
  \\
  \IDR_{10}&=G(0)\,,
  \\
  \IDR_{20}&=\int_x [G(x)]^2\,,
  \\
  \IDR_{21}&=-\int_x G(x)\ddot{G}(x)\,,
\end{align}
where we have introduced the notation 
$\ddot G(x)=\partial^2G(x)/\partial x_0^2$.

The $\IDR_{nm}$ can be obtained by taking the limit
of zero mass of the massive sums considered in the previous 
subsections. Here we will only consider the cases $m=0,1$
and for these we have\footnote{The subtraction in 
\eqref{IDRn0} is the removal of singularity due to the zero mode
for $M\to 0$. In \eqref{IDRn1} the zero mode is not singular in 
the massless limit.}
\begin{align}
  \IDR_{n0}&=\frac{1}{\Gamma(n)}\lim_{M\to0}
  \left[G_n-\frac{\Gamma(n)}{M^{2n}V_D}\right]\,,
  \label{IDRn0}
  \\
  \IDR_{n1}&=\frac{1}{\Gamma(n)}\lim_{M\to0}G_{n,1}\,.
  \label{IDRn1}
\end{align}

Now for $d=2$:
\begin{align}
  G_1&=-\frac{1}{2\pi}\ln(M)
  -\frac{1}{2\pi}\left[\frac{1}{D-2}+\Cov+\frac12\right]+g_1
  +\order{D-2}\,,
  \\
  G_2&=\frac{1}{4\pi M^2}+g_2\,,\,\,\,\,\,\,\,\,d=2\,.
\end{align}
For $d=3$:
\begin{align}
  G_1&=-\frac{M}{4\pi}+g_1\,,
  \\
  G_2&=\frac{1}{8\pi M}+g_2\,,\,\,\,\,\,\,\,\,d=3\,,
\end{align}
and for $d=4$:
\begin{align}
  G_1&=\frac{M^2}{8\pi^2}\left(\ln M+\frac{1}{D-4}+\Cov\right)
  +g_1+\order{D-4}\,,
  \\
  G_2&=-\frac{1}{8\pi^2}\left(\ln M+\frac{1}{D-4}+\Cov+\frac12\right)
  +g_2+\order{D-4}\,.
\end{align}
The behavior of $g_1,g_2$ for $M\to 0$ is readily extracted from
\eqref{grplus1} and the representations \eqref{g0d2},\eqref{g0d3}
and \eqref{g0d4}.

So for the corresponding massless sums we obtain for $d=2$:

\begin{align}
  \IDR_{10}&=-\beta_1
  -\frac{1}{2\pi}\left[\frac{1}{D-2}+\Cov-\ln L+\order{D-2}\right]
  \\
  &=-\frac{1}{2\pi}
  \left[\frac{1}{D-2}-\ln L-\frac12\alpha_1+\frac{1}{2\mcV}
    +\order{D-2} \right]\,,
  \\
  \IDR_{20}&=L^2\beta_2\,,
  \\
  \IDR_{21}&=-\frac{1}{4\pi}
  \left[\frac{1}{D-2}-\ln L-\frac12\gamma_2
    +\order{D-2} \right]\,,
  \\
  \IDR_{31}&=\frac{L^2}{64\pi^2}\left[\gamma_3+1\right]\,.
\end{align}

For $D=3$:
\begin{align}
\IDR_{10}&=-\beta_1 L^{-1}\,,
\\
\IDR_{20}&=\beta_2 L\,,
\\
\IDR_{21}&=\frac{1}{8\pi L}\left(\gamma_2-2\right)\,,
\\
&=-\frac{1}{3L}\beta_1\,\,{\rm for}\,\,\,\ell_\mu=1\,\forall \mu\,,
\\
\IDR_{31}&=\frac{L}{64\pi^2}\left[\gamma_3+2\right]\,.
\end{align}

Finally for four dimensions:
\begin{align}
  \IDR_{10}&=-\beta_1 L^{-2}\,,
  \\ 
  \IDR_{20}&=\beta_2+\frac{1}{8\pi^2}\left[\ln L-\frac{1}{D-4}
    -\Cov\right]\,,\,\,\,\,\,D\sim 4\,,\label{IDR20d4}
  \\
  &=\frac{1}{8\pi^2}\left[\ln L-\frac{1}{D-4}\right]+\gov_2
  +\order{D-4}\,,
  \\ 
  \IDR_{21}&=\frac{1}{8\pi L^2}\left(\gamma_2-1\right)\,,\label{IDR21d4}
  \\
  \IDR_{31}&=\frac{1}{32\pi^2}\left[\ln L-\frac{1}{D-4}
    +\frac12\gamma_3\right]+\order{D-4}\,,
\end{align}
where
\begin{equation}
  \gov_2=\beta_2
  -\frac{\Cov}{8\pi^2}
  =\frac{1}{16\pi^2}\left[\alpha_2-\frac{1}{2\mcV}\right]\,.
  \label{ovg2}
\end{equation}
For $\ell_1=\ell_2=\ell_3=1$ and $\ell_0=\ell$ we obtain   
\begin{alignat}{2} 
  \gov_2 & =-0.0015995623298662568192\,, \qquad &
  \text{for } \ell=1\,,\label{ovg2ell1}
  \\ 
  \gov_2&=\phm 0.0079456407220530589562\,, &
  \text{for } \ell=2\,.\label{ovg2ell2}
\end{alignat}  

For the totally symmetric volume we have for $d=4$:
\begin{alignat}{2} 
  \IDR_{21}&=\frac14\IDR_{10}=-\frac{1}{4L^2}\beta_1\,,
  & \text{\ \ for } \ell=1\,,
  \\
  \IDR_{31}&=\frac{1}{D}\IDR_{20}
  =\frac{1}{32\pi^2}\left[\ln L-\frac{1}{D-4}+\frac14\right]
  +\frac14\gov_2+\order{D-4} \,,
  & \text{\ \ for } \ell=1 \,.
\end{alignat}
So for the symmetric case we have
\begin{alignat}{2} 
  \gamma_2&=1-2\pi \beta_1\,,   & \text{for } \ell=1\,,
  \\
  \gamma_3&=16\pi^2\gov_2+\frac12\,, \qquad  & \text{for } \ell=1\,.
\end{alignat}
and this has been checked numerically. Some values of 
$\gamma_s$ are given in table~\ref{gamma123}.\footnote{Due to
the subtraction term $1/t^3$ on the interval $t\in[0,1]$ 
in \eqref{gammadef} one has $\gamma_1 \sim -1/(2\ell^{4})$ when $\ell\ll 1$.}

%
\begin{table}[ht]
  \centering
  \begin{tabular}[t]{|l|l|l|l|}
    \hline
    \,\,\,$\ell$ & \qquad\qquad\quad $\gamma_1$ & \qquad\qquad\quad $\gamma_2$
    & \qquad\qquad\quad $\gamma_3$ \\[0.5ex]
    \hline 
    3.0  & $\phmzz 1.5084066631353647$ & $\phmz 4.4458864316201319$ 
      & $\,\,23.1870844937351901$ 
      \\ 
    2.0  & $\phmzz 1.1748067357894556$
      & $\phmz 2.3510031738406602$ 
      & $\phm 6.5174542845883118$  \\ 
    1.0  & $\phmzz 0$ & \phmzz $0.1174575993893936$ 
           & $\phm 0.2474072414293639$
      \\ 
    0.5  & $\,\,\,-10.0254579887014333$ & $\,\,\,-3.1877267490551846$  
        & $-1.3395264586420596$  \\ 
    0.25 & $-167.9412483865704613$ & $-15.7551608044733235$ 
        & $-2.7263973004934400$ \\ 
    \hline                              
  \end{tabular}
  \caption{\footnotesize Values for $\gamma_1,\gamma_2,\gamma_3$ for 
    $d=4$, $\ell_1=\ell_2=\ell_3=1$ and $\ell_0=\ell$,
    for different values of $\ell$.}
  \label{gamma123}
\end{table}

\subsection{The free propagator with periodic bc in DR}
\label{DRpropagator} 

Again we consider a $D$ dimensional volume with $L_0\equiv L_t$\,,
$L_1=\dots=L_{d-1}\equiv L_s$, and $L_d=\ldots=L_{D-1}\equiv\Lhat$ 
with periodic boundary conditions in each direction. 

The goal of this section is to give the analytic 
basis for writing efficient programs for $G(x)$ 
for the numerical evaluation of classes of Feynman diagrams
in coordinate space. For this purpose we find it convenient
to write the finite volume propagator 
for the massless case, with the zero mode subtracted as
\begin{equation}
  G(x) = \Delta(x) + g(x)\,,
  \label{Gx}
\end{equation}
where $\Delta(x)$ is the infinite volume propagator
and the finite volume piece $g(x)$ will be considered in detail
in subsect.~\ref{propg}.
Note that $G(x)$ satisfies the periodic boundary conditions,
while $\Delta(x)$ and $g(x)$ do not. The singularity of $G(x)$
is given entirely by $\Delta(x)$, accordingly $g(x)$ is a smooth function.

\subsubsection{Propagator in infinite volume, $D$-dimensions} 
\label{prop_inf_vol}

For the infinite volume propagator one has
\begin{equation}
  \Delta(x) = \int_0^\infty \frac{\mrd \lambda}{(4\pi\lambda)^{D/2} }
  \mre^{-x^2/(4\lambda)} = A_D r^{2-D}\,,
  \label{Delta0}
\end{equation}
where $r=|x|$ and
\begin{equation}
  A_D=\frac{\Gamma(D/2-1)}{4\pi^{D/2}}\,.
  \label{Adef}
\end{equation}
This is related to the area of the unit sphere
\begin{equation}
  \Omega_D=\frac{2\pi^{D/2}}{\Gamma(D/2)} \,,
  \label{Omegad}
\end{equation}
\begin{equation}
  (D-2)A_D\Omega_D=1\,.
\end{equation}
The first and second ``time" derivatives over $t=x_0$
are given by
\begin{align} 
  \dot{\Delta}(x)&=-(D-2)A_D\frac{t}{r^{D}}\,,\label{dDx}
  \\
  \ddot{\Delta}(x)&=(D-2)A_D\frac{Dt^2-r^2}{r^{D+2}}\,. \label{ddDx}
\end{align}
In particular for $D=4$ one has $\Omega_4=2\pi^2$ and
\begin{align}
  \Delta(x)&=\frac{1}{4\pi^2 r^2}\,, \label{Dx4}
  \\
  \dot{\Delta}(x)&=-\frac{t}{2\pi^2 r^4}\,,\label{dDx4}
  \\
  \ddot{\Delta}(x)&=\frac{4t^2-r^2}{2\pi^2 r^6}\,.\label{ddDx4}
\end{align}
Note that by definition (by analytic continuation from $D<2$) 
in DR one has  $\Delta(0) = 0$.

\subsubsection{Calculating $g(x)$ for general $D$}
\label{propg}

We start with the representation
\begin{equation}
  G(x)=\frac{1}{V_D} \psump \mre^{ipx}
  \int_0^\infty\mrd\lambda\,\mre^{-\lambda p^2} \,.
 \label{Gx0}
\end{equation}
Then following \cite{Has90} we split the region according to
\begin{equation}
  G(x) = \int_0^\mu \frac{\mrd \lambda}{(4\pi\lambda)^{D/2} }
  \sum_v \mre^{-(x+w)^2/(4\lambda)} -\frac{\mu}{V_D} + 
  \frac{1}{V_D}\psump  \frac{\mre^{ipx-\mu p^2}}{p^2}\,,
  \label{Gx2}
\end{equation}
where we have used the Poisson summation formula in the form
\begin{equation}
  \frac{1}{V_D} \sum_p \mre^{ipx-\lambda p^2} = 
  \frac{1}{(4\pi\lambda)^{D/2}} \sum_v \mre^{-(x+w)^2/(4\lambda)}\,,
  \label{PF}
\end{equation}
where 
\begin{equation}
  w=\left(v_0L_0,\dots,v_{D-1}L_{D-1}\right)\,,\,\,\,\,\,v_\mu\in\Z\,.
\end{equation}
For $D=4$ this gives
\begin{equation}
  \begin{split}
  g(x) = & - \frac{1}{4\pi^2 x^2}\left( 1 - \mre^{-x^2/(4\mu)}\right)
  + \psum{v}\frac{1}{4\pi^2 (x+w)^2}
   \mre^{-(x+w)^2/(4\mu)} \\
  & -\frac{\mu}{V_D} + 
  \frac{1}{V_D}\psump  \frac{\mre^{ipx-\mu p^2}}{p^2} \,.
  \qquad (D=4)
  \end{split}
  \label{gx4}
\end{equation}

Taking $\mu = \alpha L^2 /(4\pi)$ with some length scale 
$L \approx L_s \approx L_t$ and $\alpha \approx 1$ 
only a few terms are needed in the sums. 
Of course, the final result does not depend on $\alpha$.

It is convenient to use the Jacobi theta function
\begin{equation}
  \begin{split}
    S(u,z) & = \sum_{n=-\infty}^\infty \mre^{-\pi u (n+z)^2}\\
    & = u^{-1/2 }\sum_{n=-\infty}^\infty \mre^{-\pi n^2/u}
    \cos(2\pi n z) \,.
  \end{split}
  \label{Suz}
\end{equation}
The first sum above converges quickly for $1 \le  u$ while the 
second for $0 < u < 1$. One has 
\begin{equation}
    S(u,z) = 
  \begin{cases} 
    u^{-1/2}  + \order{\mre^{-\pi / u}} \,, & \text{for } u\to 0\,.
    \\
    \mre^{-\pi z^2 u} + \order{\mre^{-\pi u/4}} \,, & \text{for }
    u\to\infty\,, |z|\le 1/2 \,.
  \end{cases}
\end{equation}

Further one has
\begin{equation}
  S(u,0) = S(u)\,, \qquad 
  \int_0^1 \mrd z \, S(u,z) = u^{-1/2} \,.
\end{equation}

With this one obtains (using the arbitrary scale $L$)
the following representation for $g(x)$ 
suitable for numerical evaluation: 
\begin{equation}
  \begin{aligned}
      g(x) L^{D-2} = \frac{1}{4\pi}\int_0^\infty\mrd u
      \left\{ u^{-D/2}
        \prod_{\mu=0}^{D-1} S\left( \frac{\ell_\mu^2}{u},\frac{x_\mu}{L_\mu}\right)
        -u^{-D/2}\mre^{-\pi x^2/(L^2 u)} - \frac{1}{\mcVD}
      \right\}
  \end{aligned} 
  \label{gxdSA}
\end{equation}
and
\begin{equation}
  \begin{aligned}
      g(0) L^{D-2} & = \frac{1}{4\pi}\int_0^\infty\mrd u
      \left\{
        u^{-D/2}\mcS_D\left( \frac{\ell^2}{u}\right)
        -u^{-D/2} - \frac{1}{\mcVD}
      \right\}
      \\
      & = \frac{1}{4\pi}\int_0^\infty\mrd u
      \left\{
        \frac{1}{\mcVD} \mcS_D\left( \frac{u}{\ell^2}\right)
        -u^{-D/2} - \frac{1}{\mcVD}
      \right\} \,.
  \end{aligned} 
  \label{g0dSA}
\end{equation}

In our computations we will 
need the expansion of $g(0)$ and $\partial_\nu^2 g(0)$ in $q=D-d$
up to first order:
\begin{align}
  g(0) L^{D-2} &= g_0(0)+qg_1(0) + \order{q^2}\,,
  \\
  \partial_\nu^2{g}(0) L^D &= \partial_\nu^2{g}_0(0)+
  q\partial_\nu^2{g}_1(0) + \order{q^2}\,.
\end{align}
Using \eqref{alphadef} and \eqref{rhodef} one gets for $d>2$
\begin{equation}
    g_0(0)=\frac{1}{4\pi}\left[\alpha_1(\uell)
      -\frac{2}{d-2}-\frac{1}{\mcV}\right]\,,
  \label{g0xaA}
\end{equation}

\begin{equation}
  g_1(0) = \frac{1}{4\pi} \left[ \rho_1(\uell,\ellhat)
    + \frac{2}{(d-2)^2} + \frac{\ln(\ellhat)}{\mcV} \right] \,.
  \label{g0xa1A}
\end{equation}

Some values of $g_0(0),g_1(0)$ for $d=4$ are given in table~\ref{tab:g0}.
For $d=4$, $L_t\gg L_s$ one has 
$g(0) L_s^2 =-0.2257849591 + L_t/(12 L_s)$, and $g(t,\underline{x})$
decreases exponentially fast with $t$ until
$g(t=L_t/2,\underline{0}) = -\pi^2 (L_s/L_t)^2 + L_t/(12 L_s)$.

For the double derivatives one gets similarly (cf. Appendix~\ref{AppA})
\begin{equation}
  \partial_\nu^2{g}_0(0) = -\frac12 \gamma_1(\uell) + \frac{1}{d} \,,
  \label{ddg0xaA}
\end{equation}
\begin{equation}
  \partial_\nu^2{g}_1(0) = - \tau_1(\uell,\ellhat) - \frac{1}{d^2} \,.
  \label{ddg1xaA}
\end{equation}

Some values of $\partial_\nu^2 g_0(0)\,,
\partial_\nu^2 g_1(0)$ for $d=4$ obtained from \eqref{ddg0xaA},
and \eqref{ddg1xaA} are given in tables~\ref{tab:ddg0}, \ref{tab:ddg0x}
at different $\nu$, $\ell$ and $\ellhat$.
Here $\nu=0,1,x$ denotes the time, one of the spatial coordinates, 
and one of the $D-4$ auxiliary coordinates, respectively. 
The coefficients satisfy the relations \eqref{ddg0rel0}, \eqref{ddg0rel1}.
For large $\ell$ one has $\partial_0^2g_0(0,\ell) = 
\partial_0^2g_0(0,\infty)+1/\ell$, up to an exponentially small correction.

Since $g(x)$ satisfies the relation $\Box g(x) = 1/V_D$, one obtains
for $d=4$, 
\begin{equation}
  \partial_0^2 g(0) + 3\partial_1^2 g(0) + (D-4)\partial_x^2 g(0)
  =\frac{1}{\ell\, \ellhat^{D-4}}\,.
  \label{ddg0rel}
\end{equation}
Expanding in $D-4$ this leads to
\begin{align}
  \partial_0^2 g_0(0) + 3\partial_1^2 g_0(0) & =\frac{1}{\ell}\,,
  \label{ddg0rel0} 
  \\
  \partial_0^2 g_1(0) + 3\partial_1^2 g_1(0) + \partial_x^2 g_0(0) 
  & =-\frac{\ln\ellhat}{\ell}\,. \label{ddg0rel1}
\end{align}
The values given in table~\ref{tab:ddg0}, \ref{tab:ddg0x} 
satisfy these relations.
It turns out (as expected) that $\partial_1^2 g_0(0;\ell)$
approaches the $\ell\to\infty$ limit very fast.

\begin{table}[ht]
  \centering
  \begin{tabular}{|c|c|c|c|}
    \hline 
    \bigstrut[t]
    $\ell$ &  $\ellhat$ & $g_0(0)$ & $g_1(0)$ \\
    \hline
    1  &  1 & -0.140460985545365753 &  0.04602401621995892 \\
    1  &  2 &                       &  0.17329328071528666 \\
    1  &  3 &                       &  0.38989530371165459 \\
    1  &  4 &                       &  0.69314718058471526 \\
    1  & 10 &                       &  4.33216987849965818 \\
    1  & 15 &                       &  9.74738222662423091 \\
    \hline
    2  &  1 & -0.059114936482781319 &  0.04524519886669500\\
    2  &  2 &                       &  0.09156830194292184 \\
    2  &  3 &                       &  0.19876546078660153 \\
    2  &  4 &                       &  0.35033070353224446 \\
    2  & 10 &                       &  2.16983884163206161 \\
    2  & 15 &                       &  4.87744501569431526 \\
    \hline
    3  &    &  0.0242150467817175   &  \\
    4  &    &  0.1075483739041890  &  \\
    10  &    &  0.6075483738925662  &  \\
    20  &    &  1.4408817072258595  &  \\
    \hline
  \end{tabular}
  \caption{Numerical values of $g_0(0;\ell),g_1(0;\ell,\ellhat)$
    for $d=4$ at different values of $\ell$ and $\ellhat$.
    For large $\ell$ one has $g^{(d=4)}(0;\ell)=g^{(d=3)}(0;1) + \ell/12$,
    where $g^{(d=3)}(0;\ell=1)=-0.22578495944$.}
  \label{tab:g0}
\end{table}

\begin{table}[ht]
  \centering
  \begin{tabular}{|c|c|c|c|c|}
    \hline \bigstrut[t]
    $\nu$ & $\ell$ & $\ellhat$ & $\partial_\nu^2g_0(0;\ell)$ 
    & $\partial_\nu^2g_1(0;\ell,\ellhat)$ \\
    \hline
    0  &  1  &  1 &     0.25              & -0.0625               \\
    0  &  1  &  2 &                       & -0.03163874689457073  \\
    0  &  1  &  3 &                       & -0.03157423948306426  \\
    0  &  1  & 15 &                       & -0.03157409482132951  \\
    \hline
    0  &  2  &  1 & -0.33740336789472781  & -0.10189378657955458  \\
    0  &  2  &  2 &                       &  0.08549134113439395  \\
    0  &  2  &  3 &                       &  0.09641825496707322  \\
    0  &  2  & 15 &                       &  0.09704905965741384  \\
    0  &  3  &    & -0.5042033315676696   &    \\
    0  &  4  &    & -0.5875369102375624   &    \\
    0  & 10  &    & -0.7375369106960411   &    \\
    0  & 20  &    & -0.7875369106959981   &    \\
    0  & $\infty$ & $\infty$ & -0.83753691069608 & \\
    \hline
    1  &  1  &  1 &     0.25              & -0.0625               \\
    1  &  1  &  2 &                       & -0.03163874689457073    \\
    1  &  1  &  3 &                       & -0.03157423948306426   \\
    1  &  1  & 15 &                       & -0.03157409482132951   \\
    \hline
    1  &  2  &  1 &  0.27913445596490927  & -0.05908022312845156  \\
    1  &  2  &  2 &                       & -0.03155385450654626  \\
    1  &  2  &  3 &                       & -0.03150918768108555 \\
    1  &  2  & 15 &                       & -0.03150910214106698 \\
    1  &  3  &    &    0.279178888300334  &                      \\
    1  &  4  &    &    0.279178970079188  &                      \\
    1  & $\infty$ & &  0.279178970232028  &                      \\ 
    \hline
  \end{tabular}
  \caption{Numerical values of 
    $\partial_\nu^2 g_0(0;\ell)$, $\partial_\nu^2 g_1(0;\ell,\ellhat)$
    for $\nu=0,1$. 
  }
  \label{tab:ddg0}
\end{table}

\begin{table}[ht]
  \centering
  \begin{tabular}{|c|c|c|c|c|}
    \hline \bigstrut
    $\nu$ & $\ell$ & $\ellhat$ & $\partial_\nu^2g_0(0;\ell)$ 
    & $\partial_\nu^2g_1(0;\ell,\ellhat)$ \\
    \hline
    x  &  1  &  1 &     0.25              & -0.0625          \\
    x  &  1  &  2 & -0.56659219298166237  &  0.17118801988367523  \\
    x  &  1  &  3 & -0.97231532913191441  &  0.53453238828804972   \\
    x  &  1  &  4 & -1.25998268327401261  &  0.89196426717182040  \\
    x  &  1  & 10 & -1.91028900870890396  &  2.58200729545875239  \\
    x  &  1  & 15 & -1.92086060450589695  &  3.59782618606441357  \\
    \hline
    x  &  2  &  1 &  0.27913445596490927  & -0.05908022312845156 \\
    x  &  2  &  2 & -0.33740336789472781  &  0.08549134113439395 \\
    x  &  2  &  3 & -0.55119683545590228  &  0.27446576838970181 \\
    x  &  2  &  4 & -0.69562959457825228  &  0.45372865145015078 \\
    x  &  2  & 10 & -1.02081444723100106  &  1.29878409706335203 \\
    x  &  2  & 15 & -1.02610024512982040  &  1.80669354236667127 \\
    \hline
  \end{tabular}
  \caption{Numerical values of 
    $\partial_\nu^2 g_0(0;\ell)$, $\partial_\nu^2 g_1(0;\ell,\ellhat)$
    for the second derivative taken along one of the ``extra'' dimensions.
  }
  \label{tab:ddg0x}
\end{table}

In $D=4$ one has for small $x$
\begin{equation}
  g(x) = g(0) + \frac16 \left(
    \ddot{g}(0)-\frac{1}{4\ell} \right)
  (4 t^2-x^2) + \frac{1}{8\ell} x^2 + \order{x^4}\,.
  \label{gx_exp}
\end{equation}

\subsection{Evaluation of some 1-loop integrals using
the representation in subsect.~\ref{DRpropagator}}

As an illustration of the coordinate method that we will use to compute
the sunset diagrams in sect.~\ref{sunset}, we first apply them to 
the computation of $\int_x G(x)^2$ for $d=4$.
The latter was already treated in \cite{Has90}
using the momentum-space representation and the result presented 
in \eqref{IDR20d4}.
We evaluate this Feynman graph in position-space using the decomposition
$G(x)=\Delta(x) + g(x)$ and writing the result as a sum of different 
terms, each of which is calculated using appropriate methods,
e.g. using subtraction, splitting the integration domain, etc.
Comparing the result with the momentum space result of \cite{Has90}
is useful as a test of some subroutines that were used for numerical 
evaluation of the sunset diagram treated in sect.~\ref{sunset}.

\subsubsection{$\int_x \left[G(x)\right]^2$}

\begin{figure}[ht]
\begin{center}
\psset{unit=1.7cm}
\begin{pspicture}(-1.3,-1.2)(5.3,1.1)
\psframe(-1,-1)(5,1)
\psline[linestyle=dashed,dash=0.16 0.16](1,1)(1,-1)
\psline[linestyle=dashed,dash=0.16 0.16](3,1)(3,-1)
\pscircle(2,0){0.66}
\rput(2.0,0){{\Large $S$}}
\rput(2.65,-0.73){\Large $V_0\backslash S$}
\rput(4,0){\Large $V\backslash V_0$}
\end{pspicture}
\caption{
Splitting the regions of integration for the torus
of volume 
$V=L_t\times L_s^{d_s}\times \hat{L}^q$. From $V$ one cuts out the
hypercube $V_0$ and within this a $D=d+q$ dimensional sphere $S$ 
with the radius $\rho L_s$. The singularity in $D$ appears only
in the integral over $S$, hence in the integrals over 
$V_0\backslash S$ and $V\backslash V_0$ are evaluated in $d$
dimensions. In the figure the longest direction is $L_t$,
the other $D-1$ dimensions $L_\mu$ are symbolized by the vertical extent
of the box. 
}
\end{center}
\label{figure1}
\end{figure}
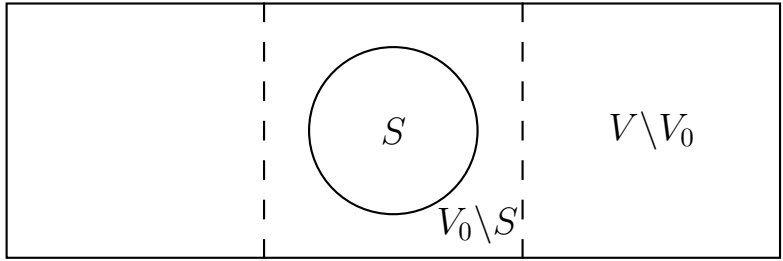

We start (taking $L_s=1$) by separating the cube 
$V_0=[-1/2,1/2]^4 \times [-\ellhat/2,\ellhat/2]^{D-4} $ 
from the total volume 
$V=[-\ell/2,\ell/2]\times[-1/2,1/2]^3\times [-\ellhat/2,\ellhat/2]^{D-4}$.
The splitting is illustrated in fig.~\ref{figure1}.
Further we decompose
\begin{equation}
  \int_V \left[G(x)\right]^2 = \sum_{r=1}^5\Sigma_r\,,
 \label{iGsq}
\end{equation}
with
\begin{alignat}{3}
  \Sigma_1 &= \int_{V_0}\left[\Delta(x)\right]^2\,, &\quad
  \Sigma_2 &= 2g(0)\int_{V_0} \Delta(x)\,,          &\quad
  \Sigma_3 &= 2 \int_{V_0} \Delta(x)\left[g(x)-g(0)\right]\,,
  \nonumber \\
  \Sigma_4 &= \int_{V_0} \left[g(x)\right]^2\,, &\quad
  \Sigma_5 &= \int_{V\backslash V_0} \left[G(x)\right]^2\,. 
\end{alignat}

To separate the divergent terms we will also need integrals over the
$d$-dimensional sphere $S$ with radius $\rho$, 
(which will be taken to be $1/2$ in the actual numerical calculations)
as well as integrals over $V_0\backslash S$.
\begin{equation}
  \Sigma_1=\Sigma_1^{(a)}+\Sigma_1^{(b)}\,,
\end{equation}
with
\begin{equation}
  \begin{aligned}
    \Sigma_1^{(a)} & =\int_{S}\left[\Delta(x)\right]^2 
    = A_D^2 \int_0^\rho \frac{\Omega_D r^{D-1}\mrd r}{r^{2D-4}} 
    = -\frac{\Gamma(D/2-1)\,\rho^{4-D}}{4\pi^{D/2}(D-2)(D-4)}  \\
    & = -\frac{1}{8\pi^2}\left[
      \frac{1}{D-4} -\ln \rho - \frac12 \left(1 + \gamma_E +\ln \pi\right)
    \right] + \ldots  \\
    & =-\frac{1}{8\pi^2(D-4)} + 0.00845810996666960998 + \order{D-4} \,, 
  \label{Psi4a1}
\end{aligned}
\end{equation}
with $\rho=1/2$. The other contribution can be computed
numerically to high precision:
\begin{equation}
 \Sigma_1^{(b)}=\int_{V_0\backslash S}\left[\Delta(x)\right]^2 =
  0.003272085168451168 \,,\,\,\,\,(\rho=\frac12\,,d=4)\,.  
  \label{Psi4a2}
\end{equation}
So adding the two contributions we obtain
\begin{equation}
  \Sigma_1=-\frac{1}{8\pi^2(D-4)} + 0.011730195135121+\order{D-4}\,.
  \label{iDsq}
\end{equation}
Numerical integration gives 
\begin{equation}
  \int_{V_0}  \Delta(x)=0.1085872819235967\,,\,\,\,\,(d=4)\,.
  \label{iD_alt}
\end{equation}
The other terms $\Sigma_3,\Sigma_4,\Sigma_5$ can also be computed 
precisely for $d=4$ for arbitrary $\ell$. 
Finally one gets (restoring the dimensions)
\begin{equation}
  \int_V \left[G(x)\right]^2 = -\frac{1}{8\pi^2}\left\{\frac{1}{D-4}
    -\ln L\right\} +\Sigma(\ell)+\order{D-4}\,.
  \label{iGsqR}
\end{equation}
Values for $\Sigma(\ell)$ and $\Sigma_i$ for $\ell=1,\ell=2$
are given in table~\ref{Sigma}. We should have
$\Sigma(\ell)=\gov_2$ with $\gov_2$ defined
in \eqref{ovg2}.
The $\ell=1$ result agrees with Hasenfratz and Leutwyler
\cite{Has90} (cf.\ their eqs.~(C.5), (A.17)) also in \eqref{ovg2ell1}
and for $\ell=2$ in \eqref{ovg2ell2}.

\begin{table}[ht]
\centering
\begin{tabular}[t]{|c|l|l|}
\hline \bigstrut[t]
integral &$\quad\quad\quad\quad\ell=1$&$\quad\quad\quad\quad\ell=2$\\[1.0ex]
\hline 
$\Sigma_2$&$-0.030504553273361740$&$-0.012838260547502574$\\
$\Sigma_3$&$\phm 0.005284317053948634$&$\phm 0.002330969017104402$\\
$\Sigma_4$&$\phm 0.01189047875442587 $&$\phm 0.002457080245220858$\\
$\Sigma_5$&$\phm 0$&$\phm 0.004265656872109587$\\
$\Sigma$&$-0.0015995623298662+$&$\phm 0.0079456407220533$\\
\hline                              
\end{tabular}
\caption{\footnotesize Values for $\Sigma_i$ and $\Sigma$ for 
$\ell=1$ and $\ell=2$.}
\label{Sigma}
\end{table}

\subsubsection{$\int_x G(x)\ddot{G}(x)$}

Next we compute $\int G \ddot{G}$ in a similar way:
\begin{equation}
  \int_V G(x) \ddot{G}(x) =\sum_{r=1}^7\Sigma'_r\,,
  \label{GddG}
\end{equation}
with
\begin{alignat}{3}
  \Sigma'_1 &= \int_{V_0} \Delta(x)\ddot{\Delta}(x)\,, &\quad
  \Sigma'_2 &= \ddot{g}(0) \int_{V_0} \Delta(x)\,, &\quad
  \Sigma'_3 &= \int_{V_0} \Delta(x) \left[\ddot{g}(x) - \ddot{g}(0)\right]\,, 
  \nonumber \\
  \Sigma'_4 &=  g(0)\int_{V_0} \ddot{\Delta}(x)\,,  &\quad
  \Sigma'_5 &= \int_{V_0} \left[g(x) - g(0)\right]\ddot{\Delta}(x)\,, &\quad
  \Sigma'_6 &= \int_{V_0} g(x) \ddot{g}(x) \,, 
  \nonumber \\
  \Sigma'_7 &= \int_{V\backslash V_0} G(x) \ddot{G}(x)\,.
\end{alignat}
Now $\Sigma'_1=0$ since it is proportional to $\Delta(0)$ 
(which is zero in DR), and 
\begin{equation}
  \Sigma'_4 =-g(0)/4+\order{D-4}\,.
\end{equation} 
The others need numerical evaluation (note the integral 
appearing in $\Sigma'_2$ is given in \eqref{iD_alt}). 
Adding the contributions we get the desired result
\begin{equation}
  \int_V G(x) \ddot{G}(x) = \frac{1}{L^2}\Sigma'(\ell)+\order{D-4}\,.
  \label{GddGR}
\end{equation}
Numerical values for $\ell=1,2$ are given in table~\ref{Sigmap}.
We should find agreement with \eqref{IDR21d4} i.e.
$\Sigma'(\ell)=-\frac{1}{8\pi}\left(\gamma_2-1\right)$.
The value for $\ell=1$ agrees with 
$\Sigma'(1)=-g(0)_{\ell=1}/4$\footnote{Note for $\ell=1$ we have 
$\Sigma'_6 = \frac14 \int_{V_0} g(x)
  = -\frac14 \int_{V_0} \Delta(x)$},
and for $\ell=2$ with the value obtained from table~\ref{gamma123}
to 10 significant digits.

\begin{table}[ht]
  \centering
  \begin{tabular}[t]{|c|l|l|}
    \hline \bigstrut[t]
    integral &$\quad\quad\quad\quad\ell=1$&$\quad\quad\quad\quad\ell=2$\\[1.0ex]
    \hline 
    $\Sigma'_2$&$\phm 0.027146820480899175$&$-0.036637714631555824279$\\
    $\Sigma'_3$&$\phm 0$&$\phm 0.004861055079412517$\\
    $\Sigma'_4$&$\phm 0.035115246386341438250$&$\phm 0.014778734120695329750$\\
    $\Sigma'_5$&$\phm 0$&$-0.01377640717436285$\\
    $\Sigma'_6$&$-0.027146820480899175$&$\phm 0.00776222484407871$\\
    $\Sigma'_7$&$\phm 0$&$-0.03074260054253895$\\
    $\Sigma'$&$\phm 0.0351152463863414$&$-0.0537547083042711$\\
    \hline                                    
  \end{tabular}
  \caption{\footnotesize Values for $\Sigma'_i$ and $\Sigma'$ for 
    $\ell=1$ and $\ell=2$.}
  \label{Sigmap}
\end{table}

\section{Massless sunset diagram with DR}
\label{sunset}

In this section we shall calculate the dimensionless 
quantity\footnote{The notation $\Wov$ was used in eq.~(3.47) of \cite{Nie15a}.}
\begin{equation}
  \Psi = - L^{2D-4}\Wov 
  =L^{2D-4} \int_V \mrd^D x G(x)^2\ddot{G}(x)\,, 
  \label{ddG_GG}
\end{equation}
at $D\sim d=4$.
Below we put $L=\Ls=1$ for simplicity.
Inserting $G(x) = \Delta(x) + g(x)$ one gets seven terms
\begin{equation}
  \Psi = \sum_{r=1}^7\Psi_r\,,
  \label{Psis}
\end{equation}
with
\begin{alignat}{3}
  \Psi_1&=\int_{V_0} g(x)^2\ddot{g}(x) \,, &\quad
  \Psi_2&=\int_{V_0} g(x)^2\ddot{\Delta}(x)\,, &\quad
  \Psi_3&=2\int_{V_0}\Delta(x)g(x)\ddot{g}(x)\,,
  \nonumber \\
  \Psi_4&=\int_{V_0}\Delta(x)^2 \ddot{g}(x)\,, &\quad
  \Psi_5&=2\int_{V_0}\Delta(x)g(x)\ddot{\Delta}(x)\,, &\quad
  \Psi_6&=\int_{V_0}\Delta(x)^2\ddot{\Delta}(x)\,,
  \nonumber \\
  \Psi_7&=\int_{V\backslash V_0} G(x)^2\ddot{G}(x)\,.
\end{alignat}
At $D=4$ only $\Psi_4$ and $\Psi_5$ have a pole, the others are finite.

\subsection{$\Psi_1$}

Numerical integration of $\Psi_1$ is simple and results
for $\ell=1,\ell=2$ are given in table~\ref{Psi}.
Note as a check at $\ell=1$ we have 
$\Psi_1=\frac14\Sigma_4=0.01189047875442588/4
= 0.002972619688606468$ which agrees to all digits.

\subsection{$\Psi_2$}

We split $\Psi_2$ in two terms 
\begin{equation} \label{Psi2}
  \Psi_2 =  \Psi_{2a}+\Psi_{2b}\,,
\end{equation}
with
\begin{align}
  \Psi_{2a}&=g(0)^2\int_{V_0} \ddot{\Delta}(x)\,,\\
  \Psi_{2b}&=\int_{V_0} \left[g(x)^2 - g(0)^2 \right]\ddot{\Delta}(x)\,.  
\end{align}
Now
\begin{equation}
  \int_{V_0}\ddot{\Delta}(x)=\frac{1}{D}\int_{V_0}\Box\Delta(x)=-\frac{1}{D}\,,
\end{equation}
hence
\begin{equation}
  \Psi_{2a}= -\frac14 g(0)^2\,.
  \label{Psi2a}
\end{equation}

\begin{figure}[ht]
\begin{center}
\psset{unit=2cm}
\begin{pspicture}(-1.1,-1.1)(1.1,1.1)
\psframe(-1,-1)(1,1)
\psline(0,0)(1,1)
\psline(0,0)(1,0.8)
\psline(0,0)(1,0.6)
\psline(0,0)(1,0.4)
\psline(0,0)(1,0.2)
\psline(0,0)(1,0.0)
\psline(0,0)(1,-0.2)
\psline(0,0)(1,-0.4)
\psline(0,0)(1,-0.6)
\psline(0,0)(1,-0.8)
\psline(0,0)(1,-1)
\psline(0.2,-0.2)(0.2,0.2)
\psline(0.4,-0.4)(0.4,0.4)
\psline(0.6,-0.6)(0.6,0.6)
\psline(0.8,-0.8)(0.8,0.8)
\rput(0.0,-1.15){\Large $x_0$}
\rput(-1.2,0.0){\Large $x_i$}
\rput(-1,-1.15){$-1/2$}
\rput( 1,-1.15){$1/2$}
\rput(-1.3,-0.9){$-1/2$}
\rput(-1.2, 0.9){$1/2$}
\end{pspicture}
\caption{
For integration over the hypercube $V_0=[-1/2,1/2]^d$ 
or over $V_0\backslash S$ it is convenient to split $V_0$ into $2d$ pyramids 
and introduce in each of them new variables $\eta$ and $u_i$ where 
$x_0=\eta$ and $x_i=\eta u_i$ (cf.~\eqref{xetau}). 
This trick improves the precision
of integration in the case when one has an integrable
singularity at $x=0$.
}
\end{center}
\label{figure2}
\end{figure}
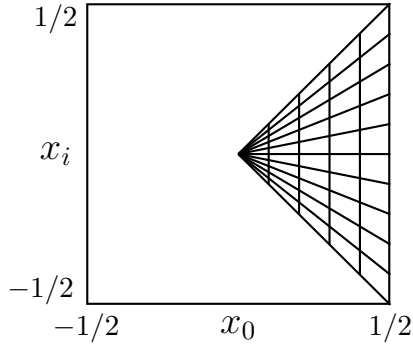

$\Psi_{2b}$ is zero for $\ell=1$.
A direct calculation for $\ell=2$ in Cartesian coordinates
is not too precise, and we get $\Psi_{2b}=0.001458\ldots \,, \qquad \ell=2$. 
The reason for poor convergence is the integrable singularity 
at the origin. 
One can improve drastically the convergence by changing the variables
\begin{equation} \label{xetau}
  x_0 = \eta \,, \quad x_i= \eta u_i\,, \quad i=1,2,3\,.
\end{equation}
The region $\eta\in[0,1/2]$, $u_i\in[-1,1]$ corresponds 
to the pyramid with the hyper-face $x_0=+1/2$ as basis.\footnote{These 
variables are also convenient to describe the part of the pyramid cut out 
by sphere $S$:
$1/\left( 2\sqrt{1+\mathbf{u}^2}\right) \le \eta \le 1/2$.}
The change of variables is illustrated in fig.~\ref{figure2}.
Precise values of $\Psi_{2a},\Psi_{2b}$ for $\ell=1,\ell=2$
are given in table~\ref{Psi}.

\subsection{$\Psi_3$}

We also split $\Psi_3$ in two parts
\begin{equation} \label{Psi3}
  \Psi_3 =  \Psi_{3a}+\Psi_{3b}\,,
\end{equation}
with
\begin{align}
 \Psi_{3a} &= 2g(0)\ddot{g}(0)\int_{V_0} \Delta(x)\,,  \label{Psi3a}
 \\
 \Psi_{3b} &=  2\int_{V_0}  \Delta(x)\left[
   g(x)\ddot{g}(x) - g(0)\ddot{g}(0) \right]\,.  \label{Psi3b}
\end{align}
All quantities appearing in $\Psi_{3a}$ are already available,
and there are no problems with numerical computation of 
$\Psi_{3b}$ for $d=4$. 
Values of $\Psi_{3a},\Psi_{3b}$ for $\ell=1,\ell=2$
are given in table~\ref{Psi}.
As a check, for $\ell=1$ one has $\Psi_{3b}=\frac14\Sigma_3$
which gives the same value as in table~\ref{Psi}.

\subsection{$\Psi_4$}

We have
\begin{equation}
  \Psi_4=\Psi_{4a}+\Psi_{4b}\,,  \label{Psi4}
\end{equation}
with
\begin{align}
  \Psi_{4a} &= \ddot{g}(0)\Sigma_1\,,\label{Psi4a}
  \\ 
  \Psi_{4b} &= \int_{V_0} \Delta(x)^2\left[\ddot{g}(x)-\ddot{g}(0)\right]\,. 
  \label{Psi4b}
\end{align}
Note that in eq.~\eqref{Psi4a} $\Sigma_1$ is given in
\eqref{iDsq} and because of the pole 
one needs $\ddot{g}_1(0)$, the $\order{D-4}$ term
in $\ddot{g}(0)$ which is given in eq.~\eqref{ddg1xaA}.

Adding values for $\Psi_{4b}$ we have 
\begin{equation}
  \Psi_4(\ell,\ellhat) = -\frac{\ddot{g}_0(0;\ell)}{8\pi^2(D-4)}
  -\frac{\ddot{g}_1(0;\ell,\ellhat)}{8\pi^2} +\Psi_{4c}\,.
  \label{Psi4R}
\end{equation}
Here $\Psi_{4c}$ is the sum of $\Psi_{4b}$ and the contribution from
the constant term in \eqref{iDsq}.
Values for $\Psi_{4b},\Psi_{4c}$ are given in table~\ref{Psi}.

\subsection{$\Psi_5$}

As for $\Sigma_1$ to separate the divergent terms we separate 
from the volume a $D$-dimensional sphere $S$ with radius $\rho$:
\begin{equation}
  \Psi_5 = \Psi_{5a}+\Psi_{5b}+\Psi_{5c} +\Psi_{5d}\,,
  \label{Psi5x}
\end{equation}
with
\begin{align}
  \Psi_{5a} &= 2g(0)\int_S\Delta(x)\ddot{\Delta}(x)\,, \label{Psi5A}
  \\ 
  \Psi_{5b} &= \sum_\nu \partial_\nu^2 g(0)\psi_{5b}^{(\nu)}\,, \label{Psi5B}
  \\
  \psi_{5b}^{(\nu)} &= \int_S x_\nu^2\Delta(x)\ddot{\Delta}(x)\,, \label{psi5Bnu}
  \\
  \Psi_{5c} &=2\int_S \hat{g}(x)\Delta(x)\ddot{\Delta}(x)\,,\,\,\,\,
  \hat{g}(x)=g(x)-g(0)
  -\frac12 \sum_\nu x_\nu^2 \partial_\nu^2 g(0)\,, \label{Psi5C}
  \\
  \Psi_{5d} &=2\int_{V_0\backslash S} \ddot{\Delta}(x) \Delta(x) g(x)\,.
  \label{Psi5D}
\end{align}
$\Psi_{5a}$ is zero since by symmetry it is proportional 
to $\Delta(0)$. 

Next for $\nu=0$ we have
\begin{multline}
  \psi_{5b}^{(0)}=
  \int_S \ddot{\Delta}(x) \Delta(x) t^2 
  =(D-2)A_D^2 \int_S \frac{(Dt^2-r^2)t^2}{r^{2D}}\\
  =(D-2)A_D^2 \Omega_D \int_0^\rho \frac{\mrd r}{r^{D-3}}
  \left\langle D \cos^4\vartheta - \cos^2 \vartheta\right\rangle \,,
\end{multline}
where $\rho=1/2$.
The averaging over $\vartheta$ is with weight $\sin^{D-2}\vartheta$.
One has
\begin{equation}
  I_D=\int_0^\pi \sin^{D-2}\vartheta \mrd \vartheta
  = \frac{\Gamma(1/2)\,\Gamma(D/2-1/2)}{\Gamma(D/2)}
\end{equation}
and $\Omega_D=\Omega_{D-1}I_D$.
For general $n$ we have
\begin{equation} \label{cosn}
  \langle \cos^n\vartheta \rangle = 
  \frac{\Gamma(n/2+1/2)\,\Gamma(D/2)}{\Gamma(n/2+D/2)\,\Gamma(1/2)}\,.
\end{equation}
The averages appearing above are
\begin{align}
  \langle \cos^2\vartheta \rangle &= \frac{1}{D}\,,
  \\
  \langle \cos^4\vartheta \rangle &= \frac{3}{D(D+2)}\,.
\end{align}
Since $\int_S \ddot{\Delta}(x) \Delta(x) r^2=0$ by symmetry, we obtain
\begin{equation}
  \psi_{5b}^{(\nu)}= \frac{2^{D-5}\Gamma(D/2-1)}{\pi^{D/2}(D-4)D(D+2)} 
  \left( 1 -D\,  \delta_{\nu 0}  \right)\,.
  \label{psi5bnu}
\end{equation}
Using eq.~\eqref{ddg0rel} one has
\begin{equation}
  \sum_\nu \partial_\nu^2 g(0) \left( 1 -D \delta_{\nu 0}\right)
  = \frac{1}{\ell\,\ellhat^{D-4}}-D \ddot{g}(0)\,. 
\end{equation}
This gives
\begin{equation}
  \Psi_{5b}= \frac{2^{D-5}\Gamma(D/2-1)}{\pi^{D/2}D(D+2)(D-4)}
  \left[ \frac{1}{\ell\, \ellhat^{D-4}}-D \ddot{g}(0)   \right]\,. 
\end{equation}
Because of the pole at $D=4$ we need here the expansion of
$\ddot{g}(0)$ to first order in $D-4$, given in
eq.~\eqref{ddg1xaA} and table~\ref{tab:ddg0}.

Expanding in $D-4$ one finally has
\begin{multline}
  \Psi_{5b}(\ell,\ellhat)=
  \frac{1}{48\pi^2}\left( \frac{1}{D-4}
    +\ln 2 - \frac12 \gamma  -\frac12 \ln\pi 
    -\frac{1}{6} \right)
  \left( \frac{1}{\ell}- 4\ddot{g}_0(0;\ell)\right)
  \\
  - \frac{1}{48\pi^2}\left(
    \frac{\ln\ellhat}{\ell}+ \frac{1}{4\ell}
    +4\ddot{g}_1(0;\ell,\ellhat) \right) 
  +\order{D-4}\,. 
  \label{Psi5by}
\end{multline}
 As a check one can verify that $\Psi_{5b}(1,1)=0$ since
$\ddot{g}_0(0;1)=1/4$ and $\ddot{g}_1(0;1,1)=-1/16$.

The last two integrals are convergent.
Note that in $d=4$ one has
\begin{equation}
  g(x) = g(0) + \frac16 \left( \ddot{g}(0)-\frac{1}{4\ell} \right) \,.
  \label{}
\end{equation}

The integral in $\Psi_{5c}$ is convergent due to the subtraction
($\hat{g}(x)=\order{r^4}$) and it is obviously zero for $\ell=1$
due to cubic symmetry.
Numerical integration with increasing precision 
indicates convergence to zero also for $\ell > 1$
although  $\hat{g}(x)$ does not have the cubic symmetry in this case.
A closer look shows that the angular integration at fixed $r$ gives zero.
This can be explained as follows. 
Since $\Box g(x)=1/\ell$ it follows from \eqref{Psi5C} that
$\hat{g}(x)$ is a harmonic function, $\Box \hat{g}(x)=0$. 
Further, since $\hat{g}(x)=\order{r^4}$ its expansion in powers of $r$
contains only spherical harmonics of order larger than two.
The angular dependence of $\ddot \Delta(x) \propto 4t^2-x^2$ 
in $\Psi_{5c}$ is given by a spherical harmonics of order two,
hence due to the orthogonality of the spherical harmonics the angular
integration indeed gives zero.

Hence we conclude that
\begin{equation} 
  \Psi_{5c} = 0 \,.
\end{equation}

Adding together $\Psi_{5}=\Psi_{5b}+\Psi_{5d}$ one has 
\begin{equation}
  \Psi_{5}(\ell,\ellhat) = 
  \frac{1}{48\pi^2(D-4)}
  \left(\frac{1}{\ell}-4\ddot{g}_0(0;\ell)\right) 
  -\frac{1}{48\pi^2} \left(
    \frac{\ln\ellhat}{\ell}+\frac{1}{4\ell}+4\ddot{g}_1(0;\ell,\ellhat)
  \right) +\psi_5+\order{D-4}\,. 
  \label{Psi5Ra} 
\end{equation}
Values of $\Psi_{5d}$ and $\psi_5$ for $\ell=1,\ell=2$ are 
given in table~\ref{Psi}, ($\Psi_{5d}=0$ for $\ell=1\,,d=4$ 
because of symmetry).

\subsection{$\Psi_6$ and $\Psi_7$}

The integral $\Psi_6$ is zero
\begin{equation}
  \Psi_6 = \int_{V_0} \Delta(x)^2\ddot{\Delta}(x)  = 
  -\frac{1}{D} \Delta(0)^2 = 0 \,.
  \label{Psi6}
\end{equation}
Also $\Psi_7=0$ for $\ell=1$ and its value for $\ell=2$ is given in
table~\ref{Psi}.

\subsection{The final result for $\Psi$}

Collecting all terms one gets\footnote{For a general shape one
should replace here $1/\ell$ by 
$1/ \prod_{\mu=0}^{d_s} \ell_\mu = 1/\mcV$.}
for the sunset diagram eq.~\eqref{ddG_GG}
\begin{equation}
  \begin{aligned}
  \Psi(\ell,\ellhat) & =
  - \frac{1}{48\pi^2(D-4)}\left( 10\, \ddot{g}(0;\ell)
    - \frac{1}{\mcVD} \right) -\frac{1}{16\pi^2}\mcWov(\ell) + \order{D-4}
  \\
  & = -\frac{1}{48\pi^2}\left[
\frac{1}{(D-4)}
  \left\{ 10\, \ddot{g}_0(0;\ell)- \frac{1}{\ell}\right\}
  +10 \, \ddot{g}_1(0;\ell,\ellhat) + \frac{\ln\ellhat}{\ell}
  +3\mcWov(\ell)\right]+ \order{D-4} \,.
  \end{aligned}
  \label{Psi_d4}
\end{equation}
The sum of non-singular terms in $\Psi(\ell,\ellhat)$ are collected
in $\mcWov(\ell)$. Its values for $\ell=1,2$ are given in table~\ref{Psi}.

According to \eqref{ddg0xaA}, \eqref{ddg1xaA} one has
\begin{align}
  \ddot{g}_0(0;\ell) & = -\frac12\gamma_1(\ell) + \frac14\,,
  \label{ddg0}
  \\
  \ddot{g}_1(0;\ell,\ellhat) & = - \mcW_1(\ell,\ellhat)
  = - \tau_1(\ell,\ellhat)-\frac{1}{16}\,.
  \label{ddg1}
\end{align}

\subsection{Checks}

For the special case $\ell=\ellhat=1$
\begin{equation}
  \int_{V_0} G(x)^2 \ddot{g}(x)
  = \frac{1}{D} \int_{V_0} G(x)^2\,.
\end{equation}
The lhs.\ is given by
\begin{equation}
  \Psi_1+ \Psi_3 + \Psi_4 =
  -\frac{1}{32 \pi^2(D-4)} + 0.00039168116473927 \,.                           
  \label{psi134}
\end{equation}
Using \eqref{iGsqR} the rhs.\ is
\begin{equation}
  \frac{1}{D} \int_{V_0} G(x)^2
  =  -\frac{1}{32\pi^2(D-4)} + 0.00039168116473921\,,
  \label{iGsqd}
\end{equation}
which agrees with \eqref{psi134} up to the 14th digit. 

The other contribution is
\begin{equation}
  \int_{V_0}  G(x)^2\ddot{\Delta}(x) = \Psi_2 + \Psi_5 + \Psi_6 \,.
  \label{psi256}
\end{equation}

Since $\Box\Delta(x)=-\delta(x)$ and using $\Delta(0)=0$ one has
\begin{equation}
  \int_{V_0} G(x)^2\ddot{\Delta}(x) 
  = -\frac{1}{D} G(0)^2 = -\frac{1}{D} g(0)^2\,,
  \label{iGsqddD}
\end{equation}
in agreement  with \eqref{psi256}.

Also we checked the results for arbitrary $\ell$ by computing 
$\Psi$ in a completely independent way outlined in appendix~\ref{AppB}.
Doing this we obtained $\mcWov(\ell=1)=0.925362611856\,,$
and $\mcWov(\ell=2)=0.154824638695$.
They differ from the results in table~\ref{Psi} in the 14th and 7th digits, 
respectively. We haven't located the source of the discrepancy
for $\ell=2$, but the estimate to 7 digits is at present sufficient for 
our purposes.

\begin{table}[ht]
\centering
\begin{tabular}[t]{|c|l|l|}
\hline \bigstrut[t]
integral &$\quad\quad\quad\quad\ell=1$&$\quad\quad\quad\quad\ell=2$\\[1.0ex]
\hline 
$\Psi_1$&$\phm 0.002972619688606468$&$-0.0002756518941414777$\\
$\Psi_{2a}$&$-0.0049323221150938617502$&$-0.0008736439288408174456$\\
$\Psi_{2b}$&$\phm 0$&$\phm 0.001457734623477771$\\
$\Psi_{3a}$&$-0.007626138318340436$&$\phm 0.004331672346637381$\\
$\Psi_{3b}$&$\phm 0.00132107926348728$&$-0.00183077295570103$\\
$\Psi_{4b}$&$\phm 0$&$\phm 0.0002857740147852$\\
$\Psi_{4c}$&$\phm 0.0029325487837802$&$-0.0036720333298669$\\
$\Psi_{5d}$&$\phm 0$&$-0.000646088213001554$\\
$\psi_5$&$\phm 0$&$-0.0019520367762068$\\
$\Psi_7$&$\phm 0$&$\phm 0.002098150204321247$\\
$\mcWov$&$\phm 0.9253626118515132$&$\phm 0.1548247146974042$\\
\hline                              
\end{tabular}
\caption{\footnotesize Values for $\Psi_i$ and $\Psi$ for 
$\ell=1$ and $\ell=2$.}
\label{Psi}
\end{table}  

Comparing the two approaches, 
note than in position-space the singularity is only at $x=0$, 
and several terms considered above have to be treated in DR to cure 
the singularity. 
In contrast to this, in the momentum-space approach the singularity 
of the sunset diagram appears in two loop-momentum variables,
hence it is more difficult to handle it.

\subsection{Other results}

$\Psi$ has no pole at $D=3$, and using similar methods
for numerical evaluation as described for $D\sim 4$
we obtain\footnote{The notation $\Wov$ was used 
in \cite{Nie15a}} (for $L=\Ls=1$, cf.~\eqref{ddG_GG} and \eqref{Psi_d4}).
\begin{equation}
  \Psi(\ell)=- \Wov(\ell) = -\frac{1}{16\pi^2}\mcWov(\ell)\,,\qquad d=3\,,
  \label{Psi_d3}
\end{equation}
with
\begin{equation}
  \mcWov(\ell)=
  \begin{cases}
    2.12506105522294\,,\qquad & \text{for } \ell=1\,,\\
    1.90198910547056\,,\qquad & \text{for } \ell=2\,.
  \end{cases}
  \label{mcWov_d3}
\end{equation}
We have also evaluated the sunset integral
\begin{equation}
  \Zov=\int_x [G(x)]^3\,,
  \label{Zov}
\end{equation}
by similar methods. 

For $D\sim 3$
\begin{equation}
  \Zov\sim\frac{1}{(4\pi)^3}
  \left\{-2\pi\left[\frac{1}{D-3}-2\ln L\right]
    +\mcZov(\uell)\right\}\,,
  \label{Zov_d3}
\end{equation}
with
\begin{equation}
  \mcZov=
  \begin{cases}
    -10.290523702796\,,\,\,\,\,& \text{for } \ell=1\,,\\
    -4.9484964492404\,,\,\,\,\,& \text{for } \ell=2\,.
  \end{cases}
\end{equation}

Next for $D\sim 4$:
\begin{equation}
  \Zov\sim\frac{1}{(4\pi)^3 L^2}
  \left\{6\left[\frac{1}{D-4}-2\ln L\right]\mathcal{Z}_0(\uell)
    +6\mathcal{Z}_1(\uell,\ellhat)
    +\mcZov(\uell)\right\}\,,
  \label{Zov_d4}
\end{equation}
where  
\begin{equation}
  \begin{aligned}
    \mathcal{Z}_0(\uell)
    &=-4\pi g_0(0;\uell)= -\alpha_1(\uell) + 1 + \frac{1}{\mcV}  \,,
    \\
    \mathcal{Z}_1(\uell,\ellhat)&=-4\pi g_1(0;\uell,\ellhat)=
   -\rho_1(\uell,\ellhat) - \frac12 - \frac{\ln(\ellhat)}{\mcV}
  \end{aligned}
\end{equation}
(cf. \eqref{g0xaA} and \eqref{g0xa1A}).

Some numerical values of $\mcZov$ are:
\begin{equation}
  \mcZov_{d=4}=
  \begin{cases}
    -2.502240295082\,,\,\,\,\,& \text{for } \ell=1\,,\\
    -3.240047780695\,,\,\,\,\,& \text{for } \ell=2\,.
  \end{cases}
\end{equation}

\section{Dimensionally regularized integrals on a strip}

In our paper \cite{Nie15a} we quote the result of a computation
of the mass gap of a massless O($n$) sigma model in 3+1 dimensions
with dimensional regularization. The computation involves
computing the 2-point function of chiral fields separated in 
the ``time" by distance $t$ in a volume\footnote{Note here $T$ is
the extent in the time direction not the temperature.}
\begin{equation}
  \Lambda=\left\{x;x_0\in[-T/2,T/2]\,,
    x_\mu\in[0,L]\,,{\rm for}\,\mu=1,\ldots,d_s\,,
    x_\mu\in[0,\Lhat]\,,{\rm for}\,\mu=d,\ldots,D-1\right\}\,,
\end{equation}
with periodic boundary conditions in the $D-1$ ``spatial" directions,
and free boundary conditions in the time direction 
(cf.~\cite{Lue91}).
The mass gap determines the exponential fall off of the 
2-point function for $t\to\infty$ (the limit $T\to\infty$ being taken first).

The free Green function $G_F(x,y)$ is determined by the 
following four properties:
\begin{align}
   \int_{x\in\Lambda}G_F(x,y)&=0\,\,\,\,;\,\,\,(y\in\overline{\Lambda})\,,
   \\
   \partial_0G_F(x,y)&=0\,\,\,\,;\,\,\,(x_0=\pm T/2\,,y\in\overline{\Lambda})\,,
   \\
   G_F(x,y)&=G_F(y,x)\,\,\,\,;\,\,\,(x,y\in\overline{\Lambda})\,,
\end{align}
where $\overline{\Lambda}$ is the interior of $\Lambda\,,$
and for $x,y\in\Lambda$ 
\begin{equation}
   -\Box G_F(x,y)=\delta^{(D)}(x-y)
   -\frac{1}{V_D} \,.
   \label{BoxGF}
\end{equation}
Here the second term is due to the subtraction of the zero mode and
\begin{equation}
   V_D=
   T\overline{V}_D\,,\,\,\,\,\,\,\,
   \overline{V}_D=L^{d-1}\Lhat^{D-d}\,.
\end{equation}

A representation of the Green function is 
\begin{equation}
   \begin{aligned}
     G_F(x,y)&=\frac{1}{\overline{V}_D}\left(
       -\frac12 |x_0-y_0| + \frac{x_0^2+y_0^2}{2T} + \frac{T}{12}
     \right)
     \\
     &+\sum_{m=-\infty}^{\infty}\Bigl\{R\left(x_0-y_0+2mT,\bfx-\bfy\right)
     +R\left(x_0+y_0+(2m+1)T,\bfx-\bfy\right)\Bigr\}\,,
   \end{aligned}
   \label{Grepd}
\end{equation}
where
\begin{equation}
   R(z)=\frac{1}{2\overline{V}_D}\sum_{\bfp\ne0}\frac{1}{\omega_\bfp}
   \mre^{-\omega_\bfp|z_0|}\mre^{i\bfp\bfz}\,.
   \label{Rzdef}
\end{equation}
Here the sum goes over 
$p_\mu=2\pi\nu_\mu/L_\mu \,,\,\,\mu=1,\dots,D-1$ 
with $\nu_\mu\in \Z$ and
\begin{equation}
   \omega_\bfp=|\bfp|\,.
\end{equation}
The function $R(z)$ is defined in \eqref{Rzdef} for 
all $z\in\R^D\backslash0\,$; 
in particular for $|z_0|\to\infty$ the function $R(z)$ 
falls exponentially. The singularity at $z=0$ is 
regularized dimensionally through an alternative representation
in terms of the Jacobi theta function \eqref{Suz}:
\begin{equation}
   R(z)=\frac{1}{4\pi L^{D-2}}\int_0^\infty\mrd u\,\frac{1}{\sqrt{u}}
   \mre^{-z_0^2\pi/(L^2u)}\left\{ u^{-(D-1)/2}
     \prod_{\mu=1}^{D-1}S\left(\frac{\ell_\mu^2}{u},\frac{z_\mu}{L\ell_\mu}\right)
     -\frac{1}{\mcVp}\right\}\,,
   \label{RzDR}
\end{equation}
where $\mcVp=\prod_{\mu=1}^{D-1}\ell_\mu$\,.

It turns out that the mass gap to third order PT 
is completely determined by the following three numbers 
involving $R$ 
: $R(0), \ddot{R}(0)$ and
\footnote{The notation $W=-\int \mrd y R(y)^2\partial_0^2 R(y)$
was used in eq.(5.13) of ref.~\cite{Nie15a}.}
\begin{equation}
   \Psiov = -L^{2D-4}W= L^{2D-4}\int \mrd y \, R(y)^2\partial_0^2 R(y)\,,
   \label{ovPsi}
\end{equation}
and the rest of this subsection deals with their numerical computation.

First from \eqref{RzDR} follows
\begin{equation}
   R(0)=\frac{1}{4\pi L^{D-2}}\int_0^\infty\mrd u\,\frac{1}{\sqrt{u}}
   \left\{u^{-(D-1)/2}
    \prod_{\mu=1}^{D-1}S\left(\frac{\ell_\mu^2}{u}\right)
     -\frac{1}{\ellhat^q}\right\}\,.
\end{equation}
For numerical evaluation and $\ell_\mu=1$ for $\mu=1,\dots,d_s$ 
and $d>2$:
\begin{equation}
     R(0)=\frac{1}{4\pi L^{d-2}}
     \left\{
       \int_0^\infty \frac{\mrd u}{\sqrt{u}}
       \left[S(u)^{d-1}\right]_\mathrm{sub}
       +\frac{2}{2-d}-2
       \right\}\,.
\end{equation}
  Using this we find:
\begin{alignat}{2}
   R(0)&=-0.3103732206\times L^{-1}\,,  & \text{for } d=3\,,
   \\
   R(0)&=-0.2257849594407580334832664917\times L^{-2}\,, &
   \quad \text{for } d=4\,.
\end{alignat}

\subsection{Calculation of the sunset diagram}

Next we turn to the numerical computation of $\Psiov$ 
for $D\sim 4$ in \eqref{ovPsi},
setting initially $L=1$ and recovering later the dependence 
on $L$.
In analogy to the computation of $\Psi$ in section 4 it is 
advantageous to first separate the infinite-volume propagator:
\begin{equation}
   R(z) = \Delta(z) + h(z)\,,
   \label{R1}
\end{equation}
with
\footnote{The function $R(z)$ is used in Peter Hasenfratz's
rotator paper, the propagator without the contribution
of the slow modes, $\mathbf p = 0$.
It is denoted there by $D^*(z)$, while our $h(z)$ by $\bar{D}^*(z)$;
cf.\ eqs.~(49) and (52) of ref.~\cite{Has09}.}
\begin{equation}
   h(z) = \frac{1}{4\pi}\int_0^\infty
   \frac{\mrd u}{\sqrt{u}} \mre^{-\pi z_0^2/u}
   \left[ \frac{1}{u^{(D-1)/2}}
     \prod_{\mu=1}^{D-1}S\left( \frac{\ell_\mu^2}{u},\frac{z_\mu}{\ell_\mu}\right)
     -\frac{1}{\mcVp}
     -\frac{\mre^{-\pi \mathbf{z}^2/u}}{u^{(D-1)/2}}
   \right]\,.
   \label{h}
\end{equation}

We have
\begin{equation}
  \Psiov = \sum_{r=1}^7\Psiov_r\,,
  \label{ovPsis}
\end{equation}
with
\begin{alignat}{3}
  \Psiov_1&=\int_{V_0} h(x)^2\ddot{h}(x) \,, &\quad
  \Psiov_2&=\int_{V_0} \ddot{\Delta}(x) h(x)^2 \,, &\quad
  \Psiov_3&=2\int_{V_0}\Delta(x)h(x)\ddot{h}(x)\,, 
  \nonumber \\
  \Psiov_4&=\int_{V_0}\Delta(x)^2 \ddot{h}(x)\,, &\quad
  \Psiov_5&=2\int_{V_0}\Delta(x) \ddot{\Delta}(x) h(x) \,, &\quad
  \Psiov_6&=\int_{V_0}\Delta(x)^2\ddot{\Delta}(x)\,, 
  \nonumber \\
  \Psiov_7&=\int_{V\backslash V_0}R(x)^2\ddot{R}(x)\,.
\end{alignat}

Although $h$ is defined for $\ell\to\infty$,
we find it convenient to use \eqref{R1} 
with a large but finite $\ell$, 
since the deviation decreases exponentially fast.
In this case
\begin{equation}
  \begin{aligned}
    h(x) & = g(x) + \frac{1}{\mcVp}\left(
      \frac12 |x_0| - \frac{x_0^2}{2\ell} - \frac{\ell}{12}\right)\,,  \\
    \ddot{h}(x) & = \ddot{g}(x) - \frac{1}{\mcV}
    +\frac{1}{\mcVp} \delta(x_0)\,,
  \end{aligned}
  \label{hx}
\end{equation}
where $\mcVp=\ellhat^{D-4} = 1 - (D-4)\ln\ellhat+\ldots$.
Note that $g(x)$ is a smooth function at $x=0$, hence \eqref{hx}
gives explicitly the singular part of $h(x)$ at the origin.

Assuming $\ell\gg 1$ and using the DR rule $\delta^{(D)}(0)=0$ one has
\begin{equation}
  \begin{aligned}
    R(0) & = h(0) = g(0) - \frac{\ell}{12} =
    -0.2257849594407580334832664917\,,
    \\
    \ddot{R}(0) & = \ddot{h}(0) = \ddot{g}(0) - \frac{1}{\ell} =
    -0.8375369106960818783868948293 \,.
  \end{aligned}
  \label{h0}
\end{equation}

\subsubsection{$\Psiov_1$}

\begin{equation}
  \Psiov_1 = \Psiov_{1a} + \Psiov_{1b}\,, 
  \label{Psi1}
\end{equation}
where
\begin{equation}
  \Psiov_{1a} =  \int_{V_0} h(x)^2
  \left[ \ddot{g}(x) -1/\ell\right]
  = -0.0119854538\,,
  \label{ovPsi1a}
\end{equation}
and
\begin{equation}
  \Psiov_{1b} =  \int_{V_0} \delta(x_0)h(x)^2
  = 0.0379616384\,.
  \label{ovPsi1b}
\end{equation}
So\footnote{We checked that it does not depend on the summation cut, 
$\alpha$ and $\ell$.
Also with numerical differentiation vs.\ the analytic formula for 
$\ddot{g}(x)$.}
\begin{equation} 
  \Psiov_{1} = 0.0259761846\,.      
\end{equation}

\subsubsection{$\Psiov_2$}

We split
\begin{equation}
  \Psiov_2 = 
  h(0)^2 \int_{V_0} \ddot{\Delta}(x) 
  + \int_{V_0} \ddot{\Delta}(x) \left[h(x)^2 - h(0)^2 \right]  \,.
  \label{ovPsi2}
\end{equation}
The first integral gives
\begin{equation}
  \Psiov_{2a}= -h(0)^2/4 = -0.0127447119774161875580368870\,.  
  \label{ovPsi2a}
\end{equation}

For the second one we divide the volume into 8 pyramids,
e.g. the one defined by \eqref{xetau}
with $\eta\in[0,1/2]$ and $u_i \in [-1,1]$.
The advantage of this is that the Jacobian, $\eta^3$ cancels
the integrable singularity at $x=0$.
Since the integrand is even in all components $x_\mu$
one can restrict the integration to $u_i\in [0, 1]$.
Thereby we obtain
\begin{equation}
  \Psiov_{2b}= -0.0176501762\,.
  \label{ovPsi2b}
\end{equation}
Hence
\begin{equation}
  \Psiov_{2}=   -0.0303948882\,.
  \label{ovPsi2x}
\end{equation}

Note that the parameterization \eqref{xetau} is also useful to calculate
integrals over $V_0 \backslash S$, taking
$1/\left( 2 \sqrt{1+u^2}\right) \le \eta \le 1/2$.
The integral over the whole torus can also be done this way.

\subsubsection{$\Psiov_3$}

It is convenient to decompose this into three terms
\begin{equation}
  \Psiov_3 = \Psiov_{3a}+\Psiov_{3b}
  +\Psiov_{3c}\,,
  \label{ovPsi3X}
\end{equation}
with
\begin{align}
  \Psiov_{3a}&=
  2 \int_{V_0} \Delta(x) \delta(x_0) \left[g(x) -\frac{\ell}{12}\right]\,,\\
  \Psiov_{3b}&=
  2h(0)\ddot{h}(0)\int_{V_0} \Delta(x)\,, \\
  \Psiov_{3c}&=
  2 \int_{V_0} \Delta(x) \left\{
    \left[ \ddot{g}(x)-\frac{1}{\ell} \right]h(x)-
    h(0)\ddot{h}(0)\right\}\,.
\end{align}
We obtain 
\begin{align}
  \Psiov_{3a} &= -0.0810043077\,, \label{ovPsi3a}
  \\
  \Psiov_{3b} &= -0.04903475008980288 \, \ddot{h}(0)
  = 0.04106841310696793\,,  \label{ovPsi3b}
  \\
  \Psiov_{3c} & = -0.017070896\,,   \label{ovPsi3c}
\end{align}
where in \eqref{ovPsi3b} we used \eqref{iD_alt}.
Collecting contributions
\begin{equation}
  \Psiov_{3} = -0.0490347500898029\, \ddot{h}(0) - 0.0980752037 
  = -0.057006791\,.
  \label{ovPsi3x}
\end{equation}

\subsubsection{$\Psiov_4$}

\begin{equation}
  \Psiov_4= \Psiov_{4a}+ \Psiov_{4b}+
  \Psiov_{4c}\,,
  \label{Psi4abc}
\end{equation} 
with 
\begin{align}
  \Psiov_{4a}&= \ddot{h}(0)\Sigma_1\,,
  \\
  \Psiov_{4b}&=\frac{1}{\mcVp}\int_{V_0}\Delta(x)^2\delta(x_0)\,,
  \\
  \Psiov_{4c}&=\int_{V_0} \Delta(x)^2
  \left[\ddot{g}(x)-\ddot{g}(0)\right]\,. 
\end{align}
Using \eqref{iDsq}
\begin{equation}
  \begin{aligned}
    \Psiov_{4a} &= 
    \left[-\frac{1}{8\pi^2(D-4)} + 0.011730195135120778 \right]\ddot{h}(0)
    \\
    & = -\frac{1}{8\pi^2(D-4)}\ddot{h}(0) -0.0098244713953312651\,.
  \end{aligned}
  \label{Psi4ax}
\end{equation}

For $\Psiov_{4b}$ we have
\begin{equation}
  \begin{aligned}
    & \frac{1}{\mcVp}\int_{S} \Delta(x)^2 \delta(x_0) =
    \frac{1}{\mcVp} A_D^2 \int_{S_{D-1}} |\mathbf{x}|^{-2D+4} = 
    \frac{1}{\mcVp} A_D^2\Omega_{D-1}
    \int_0^{1/2} \frac{r^{D-2}}{r^{2D-4}} \mrd r \\ 
    & \quad = \frac{1}{\mcVp}
    A_D^2\Omega_{D-1} \frac{2^{D-3}}{3-D} = 
    \frac{1}{\mcVp}
    \left(\frac{\Gamma(D/2-1)}{4\pi^{D/2}} \right)^2
    \frac{2\pi^{(D-1)/2}}{\Gamma((D-1)/2)}\,\frac{2^{D-3}}{3-D}  \\
    & \quad = -\frac{1}{2\pi^3} + \order{D-4}\,. 
  \end{aligned}
\end{equation}
Together with
\begin{equation} 
  \int_{V_0\backslash S} \Delta(x)^2 \delta(x_0) = 0.00272219663411\,,
\end{equation}
this gives
\begin{equation}
  \Psiov_{4b} =  -0.013403570582\,.
  \label{Psi4bx}
\end{equation}

We evaluate
\begin{equation} \label{ovPsi4c}
  \Psiov_{4c} = 0.00028615356\,.
\end{equation}

Collecting terms one has
\begin{equation}
  \begin{aligned}
    \Psiov_4 & =  \left( -\frac{1}{8\pi^2(D-4)} 
      + 0.011730195135120778\right)\ddot{h}(0) - 0.01311741702
    \\
    & = -\frac{1}{8\pi^2(D-4)} \ddot{h}(0) - 0.022941888418\,.
  \end{aligned}
  \label{Psi4x}
\end{equation}
Note that one needs here the expansion of $\ddot{h}(0)$ to $\order{D-4}$.

\subsubsection{$\Psiov_5$}

Expanding one has
\begin{equation}
  \Psiov_5 = 
  \Psiov_{5a}+\Psi_{5b} 
  -\frac{1}{\mcV} \psi_{5b}^{(0)}+\Psi_{5c}+\Psiov_{5d}
  +\Psiov_{5e}\,, 
  \label{ovPsi5y}
\end{equation}
with
\begin{align}
  \Psiov_{5a} &=2h(0)\int_S \ddot{\Delta}(x) \Delta(x)\,,\\
  \Psiov_{5d} &=
  \frac{1}{\mcVp} \int_S \ddot{\Delta}(x) \Delta(x) |x_0|\,,\\
  \Psiov_{5e} &=
  2\int_{V_0\backslash S} \Delta(x) \ddot{\Delta}(x) h(x)\,.
\end{align}
The terms $\Psi_{5b}$, $\psi_{5b}^{(0)}$ and $\Psi_{5c}$
are given by \eqref{Psi5B}, \eqref{psi5Bnu} and \eqref{Psi5C}, respectively.

The term $\Psiov_{5a}=0$ since by symmetry it is proportional 
to $\Delta(0)$.
The second and third one are logarithmically divergent and have a pole 
in $D-4$. Therefore we need here the expansion of
$\partial_\nu^2 g(0)$ to first order in $D-4$
given in eqs.~\eqref{ddg0xaA}, \eqref{ddg1xaA} and table~\ref{tab:ddg0}.
The last two integrals are convergent.

\begin{multline}
  \Psi_{5a}(\ell,\ellhat)=
  \frac{1}{48\pi^2}\left( \frac{1}{D-4}
    +\ln 2 - \frac12 \gamma  -\frac12 \ln\pi 
     -\frac{1}{6} \right)
   \left( \frac{1}{\ell}- 4\ddot{g}(0)\right) \\
 - \frac{1}{48\pi^2}\left(
   \frac{\ln\ellhat}{\ell}+ \frac{1}{4\ell}\right) 
   +\order{D-4}\,. 
   \label{Psi5a}
\end{multline}

Next from \eqref{psi5bnu}
\begin{multline}
  -\frac{1}{\mathcal{V}}\psi_{5b}^{(0)} =
  -\frac{1}{\mcV} \int_S \ddot{\Delta}(x) \Delta(x) x_0^2 =
  \frac{2^{D-5}\Gamma(D/2-1)(D-1)}{\ell\ellhat^{D-4} \pi^{D/2}D(D+2)(D-4)} = \\
   \frac{1}{16\pi^2\ell}\left( \frac{1}{D-4}
    +\ln 2 - \frac12 \gamma  -\frac12 \ln\pi - \ln\ellhat
     -\frac{1}{12} \right)
 +\order{D-4}\,. 
\end{multline}

\begin{multline}
  \Psiov_{5d}
  = \frac{1}{\mcVp}(D-2)A_D^2 \int_S \frac{(Dt^2-r^2)|t|}{r^{2D}}\\
  = \frac{1}{\mcVp}(D-2)A_D^2 \Omega_D 
  \frac{\rho^{3-D}}{3-D}
  \left\langle D \cos^3\vartheta - \cos \vartheta \right\rangle \,,
\end{multline}
where $\rho=1/2$ and the averaging is again
over $\vartheta\in[0,\pi/2]$  with weight $\sin^{D-2}\vartheta$.
Using \eqref{cosn} we have
\begin{align}
  \langle \cos\vartheta \rangle &= 
  \frac{\Gamma(D/2)}{\Gamma(D/2+1/2)\,\Gamma(1/2)}\,,\\
  \langle \cos^3\vartheta \rangle &=  
  \frac{\Gamma(D/2)}{\Gamma(D/2+3/2)\,\Gamma(1/2)} \,.
\end{align}
So for $D=4$ we obtain
\begin{align}
  \Psiov_{5d} &= - \frac{2}{5\pi^3}\,,  
  \\
  \Psiov_{5e} &= 0.00094695753 \,. \label{ovPsi5e}
\end{align}

Altogether (recalling $\Psi_{5c}=0$):
\begin{equation}
  \Psiov_{5} = \frac{1}{12\pi^2} \left[
    -\frac{1}{D-4} + \frac12 \gamma -\ln 2 + \frac12 \ln\pi + \frac16
  \right]\ddot{h}(0) -0.01195365624 \,.  
\end{equation}

\subsubsection{$\Psiov_6$ and $\Psiov_7$}

The integral $\Psiov_6=\Psi_6=0$ (see \eqref{Psi6}),
and numerical integration yields
\begin{equation}
  \Psiov_7 = 0.0000034832546\,. 
  \label{ovPsi7}
\end{equation} 

\subsubsection{Final result for $\Psiov$}

Collecting all terms one obtains the final result\footnote{Eq.~\eqref{PsiF}
  does not agree with Peter Hasenfratz's result
(63) where he has $0.029492025146$ instead of $0.0986829798$.
He also has a $\ln(L_s)$ term, but this is absent 
in the present convention due to the choice of the scale $L=L_s$.}

\begin{equation}
  \begin{aligned}
   \Psiov & = -\frac{5}{24\pi^2} \frac{1}{D-4} \ddot{h}(0)
    -0.0344802923 \, \ddot{h}(0) -0.1275614973 \\
    & = -\frac{5}{24\pi^2} \frac{1}{D-4} \ddot{h}(0) -0.0986829798 \,.
  \end{aligned}
  \label{PsiF}
\end{equation}

Note that the terms containing $\ell$ cancel and one can take here
the $\ell\to\infty$ limit. This result depends on the $\order{D-4}$
term of $\ddot{h}(0)$ which in turn depends on $\ellhat$, 
the size of the extra dimensions.
However, this dependence cancels from a physical quantity, 
like the mass gap.

\section{Finite volume momentum sums with 
  lattice regularization}
\label{lattice}

In this section we present results for certain one and two loop
momentum sums that we require for our computation of the free
energy in massless $\chi$-PT with lattice regularization \cite{Nie15a}.
We work in an asymmetric $d$-dimensional volume 
$L_0=L_t,L_\mu=L\,,\mu=1,\dots,d_s$ and periodic boundary 
in each direction. We work with the standard lattice action
so that the free propagator is given by
\begin{equation}\label{lattprop}
  G(x)=\frac{1}{V}\sum_{p\ne0}\frac{\mre^{ipx}}{\phat^2}\,,
\end{equation}
where the sum is over $p_\mu=2\pi n_\mu/L_\mu$, 
$n_\mu=0,\dots,L_\mu-1\,,V=\prod_{\mu=0}^{d_s}L_\mu$
 and $\phat_\mu=2\sin(p_\mu/2)\,$.
In many equations we will set the lattice spacing $a$ to 1.

Forward and backward difference operators are defined by
\begin{align}
  \partial_\mu f(x) &= f(x+\hat{\mu}) - f(x)\,,
  \\
  \partial_\mu^* f(x) &= f(x) - f(x-\hat{\mu})\,,
\end{align}
where $\hat{\mu}$ is the unit vector in the
$\mu$--direction, and the symmetric derivative 
\begin{equation}
  \nabla_\mu = \frac12 \left(\partial_\mu + \partial_\mu^*\right)\,.
\end{equation}
Defining the lattice Laplacian by
\begin{equation}
  \Box_\mu = \partial_\mu\partial_\mu^* \,,\,\,\,\,\,
  \Box  =\sum_\mu \Box_\mu\,,
\end{equation}
the propagator \eqref{lattprop} satisfies
\begin{equation}
  \Box G(x) = -\delta(x)+1/V\,.
\end{equation}
Some useful relations involving the lattice propagator are given
in appendix~\ref{AppC}.

\subsection{Some 1-loop momentum sums}

We define the following 1-loop momentum sums
\begin{align}
  I_{nm} & = \frac{1}{V} \psump
  \frac{\left(\phat_0^2\right)^m}{\left(\phat^2\right)^n}\,, \label{Inm}
  \\
  J_{nm} & = \frac{1}{V}\psump\frac{(\phat_0^2)^m
    \sum_\mu\phat_\mu^4}{(\phat^2)^n}\,, \label{Jnm}
  \\
  K_{nm} & =\frac{1}{V}\psump\frac{(\phat_0^2)^m
    (\sum_\mu\phat_\mu^4)^2}{(\phat^2)^n}\,,  \label{Knm}
  \\
  L_{nm} & =\frac{1}{V}\psump\frac{(\phat_0^2)^m}{(\phat^2)^n}
  \sum_{\mu\nu} \cos(p_\mu-p_\nu) \phat_\mu^2 \phat_\nu^2 \,,  \label{Lnm}
  \\
  J_{nmk} & =\frac{1}{V}\psump\frac{(\phat_0^2)^m
    \sum_\mu\phat_\mu^{2k}}{(\phat^2)^n}\,.  \label{Jnmk}
\end{align}

Some of these are related to each other e.g.
\begin{equation}
  \begin{split}
    L_{nm} & = \frac{1}{V}\psump
    \frac{(\phat_0^2)^m}{(\phat^2)^n}
        \left[ \left(\sum_\mu \phat_\mu^2 \cos p_\mu \right)^2
        + \sum_\mu  \phat_\mu^4 \sin^2 p_\mu \right]
    \\
    & = I_{n-2,m}-J_{n-1,m}+\frac14 K_{nm} + J_{nm3} - \frac14 J_{nm4}\,,
  \end{split}
  \label{Lnmeq}
\end{equation}
since $\cos p_\mu =1-\frac12 \phat_\mu^2$ and
$\sin^2 p_\mu = \phat_\mu^2 - \frac14 \phat_\mu^4$.

Note for all dimensions:
\begin{align}
  I_{00}&=1-a^d/V\,,
  \\
  I_{01}&=2\,,
  \\
  I_{02}&=6\,,
  \\
  I_{10}&=G(0)\,,\,\,\,\,d>2\,,
  \\
  I_{20}&=\sum_x G(x)^2\,.
\end{align}

We are interested in the expansion of these sums for $N=L/a\to\infty$.
As has been shown in ref.~\cite{Lue86} the 1-loop sums we consider here
have a cutoff dependence of the form
\begin{equation}
  A + B \ln N + N^{\delta-d}\sum_{s=0}^\infty c_s N^{-2s}\,,
  \label{latt_sum}
\end{equation}
where $\delta$ is determined by behavior of the summand,
$|k|^{-\delta}$ at small momenta.
The 2-loop sums, however, have a more general structure. 
To cover the different cases we use the notation\footnote{The 2-loop sum 
$W_{3a}$ for $d=2$ given in \eqref{W3a_d2} is an exception,
it has an extra $\ln^2 N/(8\pi^2)$ term.}
\begin{equation}
  X_A = \sum_r \left( X_{A;r} +  X_{A;rx} \ln N\right) N^{-r} \,.
  \label{XAr}
\end{equation}

\subsubsection{Leading terms}

In many cases the infinite volume limit $L\to\infty$ of the
sums (provided the limit exists) can be computed to arbitrary
precision using recursion relations in coordinate space.
This observation was first made by Vohwinkel and described
in detail for $d=4$ in ref.~\cite{Lue95} and later for
$d=2$ in \cite{Shi97} and for $d=3$ in \cite{Nec01}. 	
Some results for $d=4$ are given in 
tables~\ref{leadingId4}\,,\ref{leadingJKd4}.
(According to \eqref{XAr} the infinite volume limit is denoted
by $I_{nm;0}$, $J_{nm;0}$, etc.)

\begin{table}[ht]
\centering
\begin{tabular}[t]{|l|c|}
\hline \bigstrut[t]
$nm$&$I_{nm;0}$\\[1.0ex]
\hline 
$00$&$1$ \\
$10$&$0.1549333902310602140848372081$ \\
$11$&$0.25$ \\
$12$&$0.7066242375215119838793013966$ \\
$13$&$2.2930387971053356784850269783$ \\
$14$&$7.9054013578728483268787567644$ \\
$21$&$0.0387333475577650535212093020$ \\
$22$&$0.0950666097689397859151627919$ \\
$23$&$0.2935149141187831114192539608$ \\
$24$&$0.9869336725760196423390828121$ \\
$25$&$3.4667969623207723780809806095$ \\
$32$&$0.0162003867900594714029834178$ \\
$33$&$0.0436681142558825109634685570$ \\
\hline
\end{tabular}
\caption{\footnotesize Values for $I_{nm;0}$ for $d=4$.
(For notation cf. \eqref{XAr})} 
\label{leadingId4}
\end{table}

\begin{table}[ht]
\centering
\begin{tabular}[t]{|l|c|}
\hline \bigstrut[t]
$nm$&$J_{nm;0}$\\[1.0ex]
\hline 
$10$&$\phantom{1}2.8264969500860479355172055864$ \\
$11$&$\phantom{1}6$ \\
$12$&$18.5673720026100263486623960739$ \\
$21$&$\phantom{1}0.7066242375215119838793013966$   \\
$22$&$\phantom{1}2.1065867439525275504184371975$   \\
$23$&$\phantom{1}7.0275833199331741248758386402$  \\
$31$&$\phantom{1}0.0950666097689397859151627919$   \\
$32$&$\phantom{1}0.2665605938037799596982534915$   \\
\hline \bigstrut[t]
$nml$&$J_{nml;0}$\\[1.0ex]
\hline 
$213$&$\phantom{1}2.2930387971053356784850269783$  \\
$214$&$\phantom{1}7.9054013578728483268787567644$   \\
$313$&$\phantom{1}0.2935149141187831114192539608$   \\
$314$&$\phantom{1}0.9869336725760196423390828121$   \\
$323$&$\phantom{1}0.8658304227039986342356139427$   \\
$324$&$\phantom{1}2.9908252229698117808886771047$   \\
\hline \bigstrut[t]
$nm$&$K_{nm;0}$\\[1.0ex]
\hline 
$21$&$18.5673720026100263486623960739$   \\
$31$&$\phantom{1}2.1065867439525275504184371975$    \\
$32$&$\phantom{1}6.4861586895511410207353777372$ \\
\hline
\end{tabular}
\caption{\footnotesize Values for $J_{nm;0},J_{nml;0},K_{nm;0}$ for $d=4$.} 
\label{leadingJKd4}
\end{table}

\subsection{Expansion coefficients}
\label{exp_coeff}

To determine the expansion coefficients of the 1-loop sums
we have applied two methods. One is simply to compute the sums to a 
high precision for a large range of $N$ and fit the data
to the expected form, inserting the precisely known leading
term when available. Alternatively we obtain the coefficients
analytically as integrals involving the theta function \eqref{Suz},
as described in the next subsection.

\subsubsection{Lattice analogue of the theta function}

Consider the Fourier transform of $f(\theta)$
defined on the interval
$0\le \theta \le 2\pi$:
\begin{equation}
  \tilde{f}_m = \frac{1}{2\pi}	\int_0^{2\pi}
  f(\theta) \mre^{-im\theta}\,,
  \qquad f(\theta)=\sum_{m=-\infty}^{\infty}
  \tilde{f}_m \mre^{im\theta}\,.
\end{equation}


Multiplying the equation
\begin{equation}
  \sum_{m=-\infty}^{\infty} \mre^{-im N \theta}
  = \frac{2\pi}{N}\sum_{n=-\infty}^{\infty}
  \delta\left(\theta-\frac{2\pi}{N} n\right)\,.
\end{equation}
by $(2\pi)^{-1}f(\theta)$
and integrating over $\theta$ one obtains
\begin{equation}
  \frac{1}{N}\sum_{n=0}^{N-1} 
  f\left( \frac{2\pi}{N} n\right)
  = \sum_{m=-\infty}^{\infty} \tilde{f}_{mN}\,.
\end{equation}

Using this relation one obtains
\begin{equation}
  \begin{aligned}
    Q_N(z) = \frac{1}{N}\sum_{k=0}^{N-1}
    \exp\left( -z \hat{p}_k^2 \right)
    & = \mre^{-2z} \sum_{m=-\infty}^{\infty} I_{mN}(2z) 
    \\
    & = \phi_0(z) + 2 \sum_{m=1}^\infty \phi_{mN}(z)\,,
  \end{aligned}
  \label{Qnz}
\end{equation}
where $\hat{p}_k=2\sin(\pi k/N)$ and
\begin{equation} 
  \phi_{n}(z) = \mre^{-2z} I_{n}(2z)
\end{equation}
where $I_n(z)$ is the modified
Bessel function, which for integer $n$
is given by
\begin{equation}
  I_n(z) = \frac{1}{\pi}\int_0^\pi	
  \mre^{z\cos\theta}\cos(n\theta) \mrd \theta \,.
\end{equation}
For convenience of the reader we have summarized some properties 
of $I_n$ that we use in appendix~\ref{AppD}. 
For fixed $z$, $Q_N(z)$ approaches $\phi_0(z)$ exponentially fast.
The approach becomes slower with increasing $z$,
but the expansion \eqref{Inxn} shows that
even when the argument increases slower than $N^2$
one still has 
\begin{equation}
  \lim_{N\to\infty} \left(Q_N(c N^\alpha) - \phi_0(c N^\alpha)\right)
  = 0 \qquad \text{for } \alpha < 2\,,
\end{equation}
with the difference decreasing faster than any inverse
power of $N$.
This is not true for $z\propto N^2$, and for 
this case one obtains another scaling function.

We introduce the lattice counterpart of $S(x)$ by
\begin{equation}
  \begin{aligned}
    S_N(x) & = \sum_{k=0}^{N-1}
    \exp\left(- x N^2 \hat{p}_k^2/(4\pi)\right)
    \\
    & = N \mre^{-x N^2/(2\pi)}
    \sum_{m=-\infty}^{\infty}I_{mN}(x N^2/(2\pi))
    \\
    & = N Q_N\left(\frac{x N^2}{4\pi} \right)\,,
  \end{aligned}
  \label{SNdef}
\end{equation}
or equivalently
\begin{equation}
  Q_N(z) 
  =\frac{1}{N}S_N\left(\frac{4\pi z}{N^2} \right)\,.
  \label{QNSN}
\end{equation}
In the large $N$ limit one has
\begin{equation}
  \lim_{N\to\infty} S_N(x) = S(x)\,.
  \label{SNlim}
\end{equation}
Note that the $N\to\infty$ limit is not uniform in $x$. 
Since $S_N(0)=N$ while  $S(x)\sim 1/\sqrt{x}$ for small $x$,
one expects that $S_N(x)\approx S(x)$ holds for
$N\gg 1/\sqrt{x}$.

Similarly to the continuum case, the first
representation in \eqref{SNdef}
converges very fast for $x\ge 1$
while the second one for $x\le 1$.
In both cases one needs only a few terms in the
corresponding sum.
For $Q_N(z)$ in \eqref{Qnz} this corresponds 
to $z > z_0(N)$ and $z < z_0(N)$ with
$z_0(N) = N^2/(4\pi)$.

The relatively slowly convergent lattice sums $I_{nm}$ defined in
\eqref{Inm} can be calculated using $S_N(x)$. 
For $m=0$ one has
\begin{equation}
  \begin{aligned}
    I_{n0} & = \frac{1}{\Gamma(n)} \int_0^\infty \mrd z \, z^{n-1}
    \left[  \prod_{\mu} Q_{N_\mu}(z) - \frac{1}{V}\right]
    \\
    & = \left( \frac{N^2}{4\pi} \right)^{n}
    \frac{1}{\Gamma(n) V} \int_0^\infty \mrd x \, x^{n-1}
    \left[  \prod_{\mu} S_{N_\mu}\left( \frac{x}{\ell_\mu^2}\right)
      - 1\right]\,,
  \end{aligned}
  \label{In0}
\end{equation}
where $\ell_\mu=N_\mu/N$. ($N$ is arbitrary and one can choose it to be
the spatial size, $N=N_s$).

For $m>0$
\begin{equation}
  \begin{aligned}
    I_{nm} & = \frac{(-1)^m}{\Gamma(n)} \int_0^\infty \mrd z \, z^{n-1}
    Q_{N_0}^{(m)}(z) \prod_{\mu\ne 0} Q_{N_\mu}(z) 
    \\ 
    & =
    \frac{\left(N^2\right)^{n-m-d/2} (-1)^m}{(4\pi)^{n-m}\Gamma(n)\ell_0^{2m+1}}
    \int_0^\infty \mrd x \, x^{n-1}S_{N_0}^{(m)}\left( \frac{x}{\ell_0^2}\right)
    \prod_{\mu\ne 0} S_{N_\mu}\left( \frac{x}{\ell_\mu^2}\right)\,.
  \end{aligned}
  \label{Inm_rep}
\end{equation}
It is useful to split the integral and write
(for $\ell_1=\ldots=\ell_{d-1}=1$, $\ell_0=\ell$ and general $d$)
\begin{equation}
  \begin{aligned}
    I_{nm} & =
    \frac{(-1)^m}{\Gamma(n)}
    \int_0^{z_0} \mrd z \, z^{n-1}
    \left[  Q_{N_0}^{(m)}(z) Q^{d-1}_{N}(z) 
      -\frac{\delta_{m0}}{N^d \ell}\right]
    \\
    & \quad
    + \frac{\left(N^2\right)^{n-m-d/2} (-1)^m}{(4\pi)^{n-m}\Gamma(n)\ell^{2m+1}}
    \int_{x_0}^\infty \mrd x \, x^{n-1}
    \left[ S_{N_0}^{(m)}\left( \frac{x}{\ell^2}\right) S^{d-1}_{N}(x) 
      -\delta_{m0}\right]\,,
  \end{aligned}
  \label{Inm_gen}
\end{equation}
where $x_0=4\pi z_0/N^2$.

To obtain the expansion of $I_{nm}$ for large $N$
we need the asymptotic expansion of $S_N(x)$
\eqref{SN_as0} in the next subsection.

\subsubsection{Asymptotic behavior for $N\to\infty$}

As discussed before,
for $z=z_N \propto N^\alpha$ with $\alpha<2$ the correction
term $Q_N(z_N)-\phi_0(z_N)$
goes to zero exponentially fast for $N\to\infty$.

Expanding \eqref{SNdef} in $1/N^2$ one obtains an asymptotic expansion
\begin{equation}
  S_N^{\mathrm{as}}(x) = S(x) + \frac{1}{N^2}\overline{S}_1(x)
  + \frac{1}{N^4}\overline{S}_2(x) +\ldots
  \label{SN_as0}
\end{equation}
where
\begin{equation}
\begin{aligned}
  \overline{S}_1(x) & = \frac{\pi}{3} x S''(x)\,,
  \\
  \overline{S}_2(x) 
  & = \frac{\pi^2}{90} \left( 4 x S^{(3)}(x)+5 x^2 S^{(4)}(x)\right) \,,
  \\
  \overline{S}_3(x) & = \frac{\pi^3}{5670}
  \left( 18 x S^{(4)}(x)+84 x^2  S^{(5)}(x) + 35 x^3 S^{(6)}(x) \right)\,, 
  \\
  \overline{S}_4(x) & = \frac{\pi^4}{340200}
  \left( 48 x S^{(5)}(x)+ 696 x^2  S^{(6)}(x) + 840 x^3 S^{(7)}(x)
    + 175 x^4 S^{(8)}(x) \right)\,.
\end{aligned}
\end{equation}
As mentioned above the expansion \eqref{SN_as0} is not uniform.

We will also need the behavior of the corresponding terms at $x=0$:
\begin{alignat}{2}
  \overline{S}_1 & \sim \frac{\pi}{4} x^{-3/2}\,, & \qquad
  \overline{S}_2 & \sim \frac{9 \pi^2}{32}   x^{-5/2}\,,
  \nonumber \\
  \overline{S}_3 & \sim \frac{75 \pi^3}{128}    x^{-7/2}\,, & \qquad
  \overline{S}_4 & \sim \frac{3675 \pi^4}{2048} x^{-9/2}\,.
\end{alignat}

Introducing $z=y N^\alpha$ with $0<\alpha<2$ one has
\begin{equation}
  S_N^{\mathrm{as}}\left(4\pi N^{\alpha-2}y\right) 
  \sim N \phi_0\left( N^\alpha y \right)\,.
\end{equation}
This means that the singularity of 
$S_N^{\mathrm{as}}(x)$ at $x\sim 0$ matches the asymptotic
behavior of $\phi_0(z)=\mre^{-2z}I_0(2z)$
for $z\to\infty$.
The relation can be checked using the asymptotic
form \eqref{Ix0} and 
$S(x)=x^{-1/2}\left[1+\order{\mre^{-\pi/x}}\right]$.

The relation between the $x>1$ and $x<1$ regions
for $S(x)$ is given in \eqref{Sxi}. Differentiating it 
one obtains the corresponding relations between the derivatives
of $S(x)$.

At $x=x_0(N)$ the difference $S_N(x)-S(x)$
changes sign: for $x<x_0(N)$ one has
$S_N(x) < S(x)$ while for $x>x_0(N)$ one has
$S_N(x) > S(x)$. 
Interestingly, for $N\ge 8$
one has with a very good precision
$x_0(N) \approx 4\pi z_0/N^2$ where $z_0=0.06447351504$.

For $N\to\infty$, $N_0=\ell N$ in $d$-dimensions one has
for the bracket appearing in the integrand of \eqref{Inm_gen}
\begin{equation} 
  \begin{aligned}
    & S_{N_0}^{(m)}\left(\frac{x}{\ell^2}\right)S^{d-1}_N(x) -\delta_{m0}
    \\
    & \quad \sim \Psi_m(x;\ell) = 
    \Phi_{0m}(x;\ell) + \frac{1}{N^2}\Phi_{1m}(x;\ell) 
    + \frac{1}{N^4}\Phi_{2m}(x;\ell) + \ldots\,,
  \end{aligned}
\end{equation}
where 
\begin{align}
  \Phi_{0m}(x;\ell) & = 
  S^{d-1}(x) S^{(m)}\left(\frac{x}{\ell^2}\right)-\delta_{m0}\,, 
  \\
  \Phi_{1m}(x;\ell) & =
  S^{d-1}(x) \overline{S}_1^{(m)}\left(\frac{x}{\ell^2}\right)\frac{1}{\ell^2}
  + (d-1)S^{d-2}(x)\overline{S}_1(x)S^{(m)}\left(\frac{x}{\ell^2}\right)\,,
  \\
  \Phi_{2m}(x;\ell) & =
  S^{d-1}(x) \overline{S}_2^{(m)}\left(\frac{x}{\ell^2}\right)\frac{1}{\ell^4}
  + (d-1) S^{d-2}(x) \overline{S}_1(x) 
  \overline{S}_1^{(m)}\left(\frac{x}{\ell^2}\right)\frac{1}{\ell^2}
  \nonumber\\
  & \hspace{-3em}
  +(d-1) S^{d-2}(x)\overline{S}_2(x) S^{(m)}\left(\frac{x}{\ell^2}\right)
  +\frac12 (d-1)(d-2) S^{d-3}(x)\overline{S}_1^2(x) 
  S^{(m)}\left(\frac{x}{\ell^2}\right)\,.
\end{align}

Their leading singularity for $x\to 0$ is given by
\begin{align}
  \Phi_{0m}^\mathrm{sing}(x;\ell) & = 
  (-1)^m \frac{\Gamma(m+1/2)}{\Gamma(1/2)} \ell^{2m+1} x^{-(m+d/2)} \,,
  \\
  \Phi_{1m}^\mathrm{sing}(x;\ell) & =
  (-1)^m \frac{\pi}{4}\frac{\Gamma(m+1/2)}{\Gamma(1/2)}
  (2m+d) \ell^{2m+1} x^{-(m+1+d/2)}  \,,
  \\
  \Phi_{2m}^\mathrm{sing}(x;\ell) & =
  (-1)^m \frac{\pi^2}{32}\frac{\Gamma(m+1/2)}{\Gamma(1/2)}
  (12 m^2 + 20m +4dm+8d+d^2) 
  \nonumber \\
  & \qquad \times \ell^{2m+1} x^{-(m+2+d/2)} \,.
\end{align}

As illustrated in the next subsection, for the case when $I_{nm}$
has a finite $N\to\infty$ limit (i.e. $2(n-m) < d$) 
one obtains\footnote{Eq.s~\eqref{Inml}, \eqref{Inmlk} give the correct
answer for some higher coefficients even when $2(n-m) \ge d$, but these
cases need a special treatment, like for $I_{20}$ in $d=4$ discussed below.}
\begin{equation}
  I_{nm} =
  \frac{(-1)^m N^{2(n-m)-d}}{(4\pi)^{n-m}\Gamma(n) \ell^{2m+1}} 
  \int_0^\infty \mrd x \, x^{n-1} 
  \left[ \Psi_m(x;\ell)-\Psi^{\mathrm{sing}}_m(x;\ell)\right]\,.
  \label{Inml}
\end{equation}

Expanding in powers of $1/N^2$ one gets for the coefficient of $N^{-\nu}$
\begin{equation}
  I_{nm;\nu} =
  \frac{(-1)^m}{(4\pi)^{n-m}\Gamma(n) \ell^{2m+1}} 
  \int_0^\infty \mrd x \, x^{n-1} 
  \left[ \Phi_{km}(x;\ell)-\Phi^{\mathrm{sing}}_{km}(x;\ell)\right]\,,
  \label{Inmlk}
\end{equation}
where $k=0,1,\ldots$ and $\nu=2(k+m-n)-d$.

\subsubsection{Examples: $I_{10}\,,I_{21}$ for $d>2$}
\label{ex_I10}

Consider $I_{10}$ for $d>2$ in a $N_0\times N^{d_s}$ volume
($N_0=L_0/a=N\ell, N=L_s/a$):
\begin{equation}
  \begin{aligned}
    I_{10} &= \int_0^\infty \mrd z
    \left[ Q_{N\ell}(z)Q_{N}(z)^{d_s}-\frac{1}{N^d\ell}\right]
    \\
    & = \int_0^{z_0} \mrd z Q_{N\ell}(z)Q_{N}(z)^{d_s}-\frac{z_0}{N^d\ell}
    +\int_{z_0}^{\infty}\mrd z\left[Q_{N\ell}(z)Q_{N}(z)^{d_s}
      -\frac{1}{N^d\ell}\right]\,.
  \end{aligned}
  \label{I10}
\end{equation}
One can show that for $z_0 \propto N^2$ 
only the first term contributes to the constant
piece $I_{10;0}$ and one obtains\footnote{The result of numerical integration 
with MAPLE for $d=4$ agrees to 27 digits with the exact value given 
in table~\ref{leadingId4}.}
\begin{equation}
  I_{10;0} = \int_0^\infty \mrd z 
  \left[ \phi_0(z) \right]^d\,.
  \label{I10_0}
\end{equation}

To calculate the $1/N^r$ corrections
we consider the differences
\begin{equation}
  \Delta_A(N,z_0)=\int_0^{z_0}\mrd z 
  \left[ Q_{N\ell}(z)Q_N(z)^{d_s} - \phi_0(z)^d\right]\,,
  \label{I10a}
\end{equation}
and
\begin{equation}
  \Delta_B(N,z_0) = 
  \int_{z_0}^\infty \mrd z \left[ Q_{N\ell}(z)Q_N(z)^{d_s}-
    Q_{N\ell}^{\mathrm{as}}(z)Q_N^{\mathrm{as}}(z)^{d_s}\right]\,,
  \label{I10b}
\end{equation}
where
\begin{equation}
  Q_N^{\mathrm{as}}(z) = 
  \frac{1}{N} S_N^{\mathrm{as}}\left(
    \frac{4\pi z}{N^2}\right)\,.
  \label{QNas}
\end{equation}
For $z_0 = z_0(N) =c N^\alpha$, where $1 < \alpha < 2$
both integrals \eqref{I10a} and \eqref{I10b}
go to zero for $N\to\infty$ exponentially fast.

We have
\begin{equation}
  \begin{aligned}
    I_{10} -I_{10;0} &= 
    - \int_{z_0}^\infty \mrd z\,\phi_0^d(z)
    -\frac{z_0}{N^d\ell} 
    +\int_{z_0}^\infty \mrd z
    \left[
      Q_{N\ell}^{\mathrm{as}}(z)Q_N^{\mathrm{as}}(z)^{d_s}
      -\frac{1}{N^d\ell}\right]
    \\
    & \qquad
    +\Delta_A(N,z_0)+\Delta_B(N,z_0)\,.
  \end{aligned}
  \label{I1c}
\end{equation}
Here the $z_0$ dependence should cancel, i.e.
$Q_N^{\mathrm{as}}(z_0)$ and $\phi_0(z_0)$ should
have the same asymptotic behavior for 
$z_0(N) = c N^\alpha$ and $N\to\infty$.
Note that the argument 
$x_0(N)=4\pi z_0(N)/N^2 \propto N^{\alpha-2}$ 
of the functions $S(x)$ in \eqref{QNas} goes to
zero in this limit. Hence the contributions
from the large-$z$ asymptotic of $\phi_0(z)$ 
and the small-$x$ asymptotic of 
$S_N^{\mathrm{as}}(x)/N$ cancel each other.
This is indeed the case, one has\footnote{The $N$-dependence 
on $Q_N^{\mathrm{as}}(z)$ cancels in the asymptotic expansion.} 
\begin{equation} 
  \frac{1}{N} S_N^{\mathrm{as}}
  \left(\frac{4\pi z_0}{N^2}\right) 
  \sim \phi_0(z_0) \sim 
  \frac{1}{\sqrt{4\pi z_0}}\left(
    1 + \frac{1}{16 z_0}+ \frac{9}{512 z_0^2}
    +\ldots
  \right)\,.
\end{equation}

With $x_0=4 \pi z_0/N^2$ one has
\begin{equation}
  \begin{aligned}
    & \int_{z_0}^\infty \mrd z
    \left[
      Q_{N\ell}^{\mathrm{as}}(z)Q_N^{\mathrm{as}}(z)^{d_s}	
      -\frac{1}{N^d\ell}\right] 
    =\frac{1}{4\pi N^{d-2}\ell}
    \int_{x_0}^\infty \mrd x
    \left[
      S_{N\ell}^{\mathrm{as}}\left(\frac{x}{\ell^2}\right)
      S_N^{\mathrm{as}}(x)^{d_s}
      -1\right] \\
    & =\frac{1}{4\pi N^{d-2}\ell} 
    \int_{x_0}^\infty \mrd x \left\{\Phi_{00}(x;\ell)
      +\frac{1}{N^2}\Phi_{10}(x;\ell)
      +\frac{1}{N^4}\Phi_{20}(x;\ell)+\dots\right\}\,.
  \end{aligned}
\end{equation}

The subtraction of the integral of $\phi_0(z)^d$
amounts to subtracting from each term its singular 
part for $x\to 0$. So finally we have for $d>2$
\begin{equation} 
  I_{10}= I_{10;0} + \frac{1}{N^{d-2}}I_{10;d-2}
  + \frac{1}{N^d}I_{10;d} + \frac{1}{N^{d+2}} I_{10;d+2}+\ldots\,,
  \label{I10_ex}
\end{equation}
where

\begin{align}
  I_{10;d-2} &= \frac{1}{4\pi\ell}\int_0^\infty \mrd x\,
  \left[\Phi_{00}(x;\ell)-\frac{\ell}{x^{d/2}}\right]\,,   \label{I10_1}
  \\
  I_{10;d} &=\frac{1}{4\pi\ell}\int_0^\infty \mrd x\,
  \left[\Phi_{10}(x;\ell)-\frac{\pi d \ell}{4x^{d/2+1}}\right]\,, \label{I10_2}
  \\
  I_{10;d+2} &=\frac{1}{4\pi\ell}\int_0^\infty \mrd x\, 
  \left[\Phi_{20}(x;\ell)-\frac{\pi^2\ell d(d+8)}{32 x^{d/2+2}}\right]\,.
  \label{I10_3}
\end{align}

Comparing \eqref{I10_1} with $\beta_1$ given in \eqref{betand3}
and \eqref{betand4} for $d=3,4$ we obtain 
\begin{equation}
  I_{10;d-2}=-\beta_1\,,\qquad \text{for } d=3,4\,,
\end{equation}
relating coefficients of the lattice expansion to shape 
coefficients in DR. This is just one example of many such relations.

Repeating the steps used above one gets for the expansion
coefficients of $I_{21}$: 
\begin{equation}
  I_{21} = \frac{1}{d} I_{10;0}+\frac{1}{N^{d-2}}I_{21;d-2} +
  \frac{1}{N^d}I_{21;d}+\ldots\,,\qquad d>2\,,
  \label{I21_0}
\end{equation}
with
\begin{equation}
  I_{21;d-2} =-\frac{1}{4\pi\ell^3}\int_{0}^\infty \mrd x\, x
  \left[\Phi_{01}(x;\ell) + \frac{\ell^3}{2x^{(d+2)/2}}\right]\,.
  \label{I21_1}
\end{equation}
Again, for $d=3,4$ 
$I_{21;d-2}$ is related to the corresponding DR shape coefficient e.g. 
\begin{equation}
  I_{21;2} = \frac{1}{8\pi}(\gamma_2-1)=L^2\IDR_{21}\,,\qquad d=4\,.
\end{equation}

Next
\begin{equation}
  I_{21;d} =- \frac{1}{4\pi\ell^3}\int_{0}^\infty \mrd x\, x
  \left[\Phi_{11}(x;\ell) + \frac{(d+2)\pi\ell^3}{8 x^{(d+4)/2}}\right]\,.
  \label{I21_2}
\end{equation}

So far in this subsection we have only considered sums
which have a finite infinite volume limit. 
As an example of a sum which does not have this property we
consider $I_{20}$. For $d=3\,\,\,I_{20}$  
is linearly divergent (see \eqref{Iexp_d3ell1}\,,\eqref{Iexp_d3ell2}). 

In the rest of this subsection we only consider the case $d=4$ 
for which $I_{20}$ is logarithmically divergent for $N\to\infty$.
Here we will separate the $\sim \log(N)$ and the constant 
terms\footnote{For our work in ref.~\cite{Nie15a} 
the constant term is actually needed only for the renormalization.}.
Restricting also to $\ell=1$ we have
\begin{equation}
  I_{20} = \int_0^{z_0} \mrd z \, z \left( Q_N^4(z) - \frac{1}{N^4}\right)
  + \frac{1}{16\pi^2}\int_{x_0}^{\infty} \mrd x \, x \left( S_N^4(x) - 1\right) 
  \label{I20sol}
\end{equation}
where $x_0=4\pi z_0/N^2$. Choosing $z_0 = c N^{2-\epsilon}$ with some fixed small
$\epsilon>0$ in the first term one could replace $Q_N(z)$ by $\phi_0(z)$
up to exponentially small corrections.

One has
\begin{equation} 
  \begin{aligned} 
    I_{20}^A & = \int_0^{z_0} \mrd z \, z \left( Q_N^4(z) - \frac{1}{N^4}\right)
    \sim \int_0^{z_0} \mrd z \, z \phi_0^4(z) - \frac{z_0^2}{2 N^4}
    \\
    & =  \int_0^1 \mrd z \, z \phi_0^4(z) + 
    \int_1^{z_0} \mrd z \, z \left(\phi_0^4(z)-\frac{1}{16\pi^2 z^2}\right)
    \\ & \quad
    + \frac{1}{16\pi^2} \log z_0 - \frac{z_0^2}{2 N^4}\,.
  \end{aligned}
  \label{I20A}
\end{equation}
In the second integral of \eqref{I20sol} one can replace 
$S_N(x)$ by $S(x)$ up to $\order{1/N^2}$ correction.
\begin{equation} 
  \begin{aligned} 
    I_{20}^B & = 
    \frac{1}{16\pi^2}\int_{x_0}^{\infty} \mrd x \, x \left( S_N^4(x) - 1\right) 
    \sim \frac{1}{16\pi^2}\int_{x_0}^{\infty} \mrd x \, x \left( S^4(x) - 1\right) 
    \\ 
    & = \frac{1}{16\pi^2}\int_{x_0}^1 \mrd x \, x 
    \left( S^4(x) - \frac{1}{x^2}\right)
    + \frac{1}{16\pi^2}\int_1^\infty \mrd x \, x \left( S^4(x) - 1\right)
    \\
    & \quad - \frac{1-x_0^2}{32 \pi^2} - \frac{1}{16\pi^2} \log x_0\,.
  \end{aligned}
  \label{I20B}
\end{equation}

Adding the two terms and for large $N$ one obtains
\begin{equation} 
  \begin{aligned} 
    I_{20} & =  \int_0^1 \mrd z \, z \phi_0^4(z) + 
    \int_0^\infty \mrd z \, z \left(\phi_0^4(z)-\frac{1}{16\pi^2 z^2}\right)
    \\
    & \quad 
    +\frac{1}{16\pi^2}\int_0^1 \mrd x \, x \left( S^4(x) - \frac{1}{x^2}\right)
    + \frac{1}{16\pi^2}\int_1^\infty \mrd x \, x \left( S^4(x) - 1\right)
    \\ & \quad
    -\frac{1}{32 \pi^2} + \frac{1}{16\pi^2}\log\left(\frac{N^2}{4\pi} \right)
    + \order{1/N^2}\,.
  \end{aligned}
\end{equation}
Evaluating the integrals one has
\begin{equation} 
I_{20} =  \frac{1}{8\pi^2} \log N + 0.01004098140549470847620108 
+ \order{1/N^2}\,.
\end{equation}
The coefficients $I_{20;2}$ and $I_{20;4}$ for $d=4$ are
given in the next subsection.

One can repeat these steps for the general case of $I_{nm}$. 
Using the representation \eqref{Inm_gen} one reproduces the form 
of the expansion given by \eqref{latt_sum}.
The $\log N$ term comes from the $1/x$ and $1/z$ terms of the 
corresponding integrands while in the rest one can set 
$x_0=0$ and $z_0=\infty$, as in \eqref{I20A} and \eqref{I20B}. 
Finally, the coefficient $N^{2(n-m)-d}$ in front of the second
integral in \eqref{Inm_gen} reproduces the prefactor $N^{\delta-d}$
of the sum in \eqref{latt_sum}.

\subsubsection{Expansions for $d=4$}

We are interested in the expansion of the lattice sums for $N^{-1}=a/L\to 0$
(at a fixed aspect ratio $\ell$) up to and including
the $\order{a^4/L^4}$ terms. For the sums we require we have:
\begin{equation}
  \begin{split}
    & I_{00} = 1 - \ell^{-1}\,N^{-4} \\
    & I_{10}= I_{10;0} + I_{10;2}\,N^{-2} + I_{10;4}\,N^{-4} + \ldots \\
    & I_{11} = I_{11;0} + I_{11;4} \,N^{-4}\,, \\
    & I_{20} = \ln N /(8\pi^2) + I_{20;0} + I_{20;2}\,N^{-2} + I_{20;4}\,N^{-4}  
    + \ldots \\
    & I_{21} = I_{21;0} + I_{21;2}\,N^{-2} + I_{21;4}\,N^{-4} + \ldots \\
    & I_{22} = I_{22;0}  + I_{22;4}\,N^{-4} + \ldots \\
    & I_{31} = \ln N /(32\pi^2) + I_{31;0} + I_{31;2}\,N^{-2} 
             + I_{31;4}\,N^{-4} + \ldots \\
    & I_{32} = I_{32;0} + I_{32;2}\,N^{-2} + I_{32;4}\,N^{-4} + \ldots \\
    & I_{33} = I_{33;0} + I_{33;4} \,N^{-4} +\ldots \\
    & J_{31} = J_{31;0} + J_{31;4} \,N^{-4} +\ldots \\
  \end{split}
\end{equation}

The coefficients $I_{nm;\nu}$ can be calculated (at least for the cases
with finite infinite-volume limit) from \eqref{Inmlk}.

Next consider the lattice sum $J_{31}$ (see \eqref{Jnm});
for $d=4$ one has
\begin{equation}
  J_{31} = \frac{1}{V} \psump \frac{(\phat_0^2)^3  }{(\phat^2)^3}+ J_{31}^B\,,
  \label{J31a}
\end{equation}
with
\begin{equation}
  J_{31}^B = \frac{3}{V} \psump \frac{\phat_0^2 (\phat_1^2)^2 }{(\phat^2)^3}\,.
\end{equation}
The first term in \eqref{J31a} is $I_{33}$, hence 
\begin{equation}
  J_{31;4} = I_{33;4} + J_{31;4}^B\,,\,\,\,\,{\rm for}\,\,\,d=4\,,
\end{equation}
with
\begin{equation}
  J_{31;4}^B = - \frac{3}{2\ell^3}\int_0^\infty \mrd x \, x^2
  \left[S(x)^2 S''(x) S'\left(\frac{x}{\ell^2}\right)+\frac{3\ell^3}{8 x^5}
  \right]\,. 
  \label{J31_2lB}
\end{equation}

In tables~\ref{Ixyell123} and \ref{Ixyell45} we give values
of the coefficients above for $\ell=1,2,3$ and $\ell=4,5$ 
respectively using the integral representation \eqref{Inmlk} with MAPLE.
We checked that all results agreed to at least 12 digits in all cases with 
fits of the data (using the precise infinite volume results when available).

\begin{table}[ht]
  \centering
  \begin{tabular}{|c|l|l|l|}
    \hline \bigstrut
    & $\quad\quad\quad\quad\ell=1$ &   $\quad\quad\quad\quad\ell=2$ & 
    $\quad\quad\quad\quad\ell=3$ \\ 
    \hline
    $I_{10;2}$ & $-0.1404609855453658$ & $-0.0591149364827813$ &
    $0.0242150467817181$ \\
    $I_{10;4}$ & $\phm 0.1418858055568331$& 
    $\phm 0.1445281475197173$& $0.1582885920258339$\\
    $I_{11;4}$ & $-0.25$ & $\phm 0.3374033678947278$ & 
    $0.5042033315676824$ \\
    $I_{20;2}$ & $\phm 0.0014757515175671$ & 
    $\phm 0.0122627013456392$ & $0.0261308665866801$ \\
    $I_{20;4}$ & $\phm 0.0731806946434512$ & 
    $\phm 0.0593077967014635$ & $0.0615529993132684$ \\
    $I_{21;2}$ & $-0.0351152463863414$ & 
    $\phm 0.0537547083123951$ &  $0.1371074647314091$ \\
    $I_{21;4}$ & $\phm 0.0354714513892083$ & 
    $\phm 0.1554316713871277$ & $0.1699444287054589$ \\
    $I_{22;4}$ & $\phm 0.4256574166704993$ & 
    $\phm 0.7569495889658734$ & $0.9229739816883220$ \\
    $I_{31;2}$ & $\phm 0.0003689378793918$ & 
    $\phm 0.0167916533065880$ & $0.0307265836535553$ \\
    $I_{31;4}$ & $\phm 0.0182951736608628$ & 
    $\phm 0.0404550162559959$ & $0.0435488531765849$ \\
    $I_{32;2}$ & $\phm 0.0022136272763506$ & 
    $\phm 0.0820388915996510$ & $0.1653308457382867$ \\
    $I_{32;4}$ & $\phm 0.1647046454257431$ & 
    $\phm 0.1861803595341033$ & $0.1990444597971972$ \\
    $I_{33;4}$ & $\phm 0.5483743850461059$ & 
    $\phm 1.0693538988138826$ & $1.2370421989724505$ \\
    $J_{31;4}$ & $\phm 0.4256574166704993$ & 
    $\phm 1.3110648228057027$ & $1.4811535630769142$ \\
    \hline
  \end{tabular}
  \caption{\footnotesize Values for $I_{mn;2},I_{mn;4},J_{31;4}$ for $d=4$
    and $\ell=1,2,3$.} 
  \label{Ixyell123}
\end{table}


\begin{table}[ht]
  \centering
  \begin{tabular}{|c|l|l|}
    \hline \bigstrut[t]
    & $\quad\quad\quad\quad\ell=4$ &   $\quad\quad\quad\quad\ell=5$ \\ 
    \hline
    $I_{10;2}$ & $0.1075483739041892$ & $0.1908817072259303$ \\
    $I_{10;4}$ & $0.1652326717601757$ & $0.1693993375068985$ \\
    $I_{11;4}$ & $0.5875369102375629$ & $0.6375369106952257$ \\
    $I_{20;2}$ & $0.0400196677374950$ & $0.0539085563322884$ \\
    $I_{20;4}$ & $0.0628240364390694$ & $0.0635879153141902$ \\
    $I_{21;2}$ & $0.2204408534728138$ & $0.3037741869459578$ \\
    $I_{21;4}$ & $0.1768917723653977$ & $0.1810584492251718$ \\
    $I_{22;4}$ & $1.0063053711184669$ & $1.0563053660562876$ \\
    $I_{31;2}$ & $0.0446158019765458$ & $0.0585046924277057$ \\
    $I_{31;4}$ & $0.0448289656504658$ & $0.0455929029967931$ \\
    $I_{32;2}$ & $0.2486639742820079$ & $0.3319973068786403$ \\
    $I_{32;4}$ & $0.2059796377872626$ & $0.2101462558490117$ \\
    $I_{33;4}$ & $1.3203816834539149$ & $1.3703817074781796$ \\
    $J_{31;4}$ & $1.5645033197516197$ & $1.6145033783791748$ \\
    \hline
  \end{tabular}
  \caption{\footnotesize Values for $I_{mn;2},I_{mn;4},J_{31;4}$ for $d=4$
    and $\ell=4,5$.} 
  \label{Ixyell45}
\end{table}

\subsubsection{Expansions for $d=3$}
\label{d3exp}

Some infinite volume values are
\begin{equation}
  \begin{split}
    & I_{10;0} = 0.252731009859\,,\\
    & I_{12;0} = 0.913649942701\,,\\
    & I_{21;0} = 0.084243669953\,,\\
    & I_{22;0} = 0.164845993428\,,\\
    & I_{23;0} = 0.486683859112\,,\\
    & I_{32;0} = 0.046176554504\,,\\
    & I_{33;0} = 0.096821201585\,.\\
  \end{split}
\end{equation}

Also in this case we determined the coefficients both by fitting the 
$N$-dependence and by direct calculation using \eqref{Inmlk},
when it was applicable. In all cases we got an agreement within
the precision of the fitting procedure. (The worst case was that of
$I_{30;1}$ where the fit gave only four significant digits.)

For $\ell=1$:
\begin{align}
     & I_{10} = I_{10;0} -0.225784959441\,N^{-1}
    + 0.0428997562958 \,N^{-3}+\dots \,,
    \displaybreak[0] \nonumber \\
    & I_{11} = \frac13 -\frac13\,N^{-3}  \,,
    \displaybreak[0] \nonumber \\
    & I_{12} = I_{12;0} + \dots  \,,
    \displaybreak[0] \nonumber \\
     & I_{20} = 0.010607528892\,N + 0.012164158583
    -0.0155358881130\,N^{-1} + \ldots \,,
    \displaybreak[0]\nonumber \\
    & I_{21} = I_{21;0} -0.0752616531469\,N^{-1}
    + 0.014299918765276429\,N^{-3}+\dots  \,,
    \displaybreak[0] \nonumber \\
    & I_{22} = I_{22;0} + 0.171599025183\,N^{-3}+\dots \,,
    \displaybreak[0]                   \label{Iexp_d3ell1} \\
    & I_{23} = I_{23;0} + \dots  \,,
    \displaybreak[0]\nonumber \\
    & I_{30} = 0.000136552463\,N^3 + 0.00191507947957\,N
    + 0.000837762  \nonumber \\
    & \qquad -0.000388292243730\,N^{-1} + \ldots \,, 
    \displaybreak[0] \nonumber \\
    & I_{31} = 0.00353584296399\,N + 0.004054719528
    -0.00517862937101\,N^{-1} + \ldots \,,
    \displaybreak[0] \nonumber \\
    & I_{32} = I_{32;0} -0.0310717762260\,N^{-1}
    + 0.0973259651475\,N^{-3}+\dots  \,,
    \displaybreak[0] \nonumber \\
    & I_{33} = I_{33;0} + 0.348442731224\,N^{-3}+\dots \,.
    \nonumber
\end{align}

For $\ell=2$:
\begin{align}
    & I_{10} = I_{10;0} -0.143704325288\,N^{-1}
    + 0.0614336837790 \,N^{-3}+\dots \,,
    \displaybreak[0] \nonumber \\
    & I_{11} = \frac13 +0.218784444721 \,N^{-3}+\dots \,,
    \displaybreak[0] \nonumber \\
    & I_{12} = I_{12;0} + \dots  \,,
    \displaybreak[0] \nonumber \\
    & I_{20} = 0.020216123622\,N + 0.012164158583
    -0.00356885724612\,N^{-1} + \ldots \,,
    \displaybreak[0] \nonumber \\
    & I_{21} = I_{21;0} + 0.0114671446126\,N^{-1}
    + 0.0992873225148\,N^{-3}+\dots \,,
    \displaybreak[0] \nonumber \\
    & I_{22} = I_{22;0} + 0.578733423280\,N^{-3}+\dots \,,
    \displaybreak[0] \label{Iexp_d3ell2} \\
    & I_{23} = I_{23;0} +\dots  \,,
    \displaybreak[0] \nonumber \\
    & I_{30} = 0.001155699930\,N^3 + 0.00424327053578\,N
    + 0.000837763  \nonumber \\
    & \qquad + 0.00161399011894\,N^{-1} + \ldots \,,
    \displaybreak[0] \nonumber \\
    & I_{31} = 0.0133862498578 N + 0.004054719527
    +0.0102078527937\,N^{-1} + \ldots \,,
    \displaybreak[0] \nonumber \\
    & I_{32} = I_{32;0} + 0.0503041764078\,N^{-1}
       + 0.132427332872\,N^{-3}+\dots \,,
    \displaybreak[0] \nonumber \\
    & I_{33} = I_{33;0} + 0.847216528219\,N^{-3}+\dots \,.
    \nonumber
\end{align}

Here only $I_{20}$, $I_{30}$ and $I_{31}$ diverge for $N\to\infty$.
However, also in these cases the coefficients 
$I_{20;1}$, $I_{30;-1}$, $I_{30;1}$, $I_{31;-1}$, $I_{31;1}$ are correctly
given by \eqref{Inmlk}.

The shape coefficients $I_{10;1}$ and $I_{20;-1}$ are related to
$\beta_n$ defined in \eqref{betand3} through: 
\begin{align}
    I_{10;1} &= -\beta_1\,, \label{beta_psi_a}
    \\
    I_{20;-1} &= \phm \beta_2\,.
\end{align}
Further there are relations to $\widetilde{\beta}_n$ defined in \cite{Has93}:
\begin{align} 
    I_{21;1} &= -\frac16\widetilde{\beta}_1\,,\\                     
    I_{31;-1} &= \phm \frac{1}{12}\widetilde{\beta}_2\,.   
\end{align}

\subsubsection{Expansions for $d=2$}

In this subsection we consider only the $\ell=1$ case and
concentrate on terms which do not vanish in the infinite volume limit.

For the logarithmically divergent sum $I_{10}$ we have
\begin{equation}
  I_{10} = \int_0^{z_0} \mrd z \, \left( Q_N^2(z) - \frac{1}{N^2}\right)
  + \frac{1}{4\pi}\int_{x_0}^{\infty} \mrd x \left( S_N^2(x) - 1\right)\,, 
  \label{I10_d2}
\end{equation}
where $x_0=4\pi z_0/N^2$. Choosing $z_0 = c N^{2-\epsilon}$ with some fixed small
$\epsilon>0$ in the first term one could replace $Q_N(z)$ by $\phi_0(z)$
up to exponentially small corrections.
One has
\begin{equation} 
  \begin{aligned} 
    I_{10}^A & = \int_0^{z_0} \mrd z \left( Q_N^2(z) - \frac{1}{N^2}\right)
    \sim \int_0^{z_0} \mrd z \phi_0^2(z) - \frac{z_0}{N^2}
    \\
    & =  \int_0^1 \mrd z \phi_0^2(z) + 
    \int_1^{z_0} \mrd z \left(\phi_0^2(z)-\frac{1}{4\pi z}\right)
    + \frac{1}{4\pi} \log z_0 - \frac{z_0}{N^2}\,.
  \end{aligned}
\end{equation}
In the second integral of \eqref{I10_d2} one can replace $S_N(x)$ by $S(x)$ 
up to $\order{1/N^2}$ correction.
\begin{equation} 
  \begin{aligned} 
    I_{10}^B & = 
    \frac{1}{4\pi}\int_{x_0}^{\infty} \mrd x \left( S_N^2(x) - 1\right) 
    \sim \frac{1}{4\pi}\int_{x_0}^{\infty} \mrd x \left( S^2(x) - 1\right) 
    \\ 
    & = \frac{1}{4\pi}\int_{x_0}^1 \mrd x \left( S^2(x) - \frac{1}{x}\right)
    + \frac{1}{4\pi}\int_1^\infty \mrd x \left( S^2(x) - 1\right)
    \\
    & \quad - \frac{1-x_0}{4 \pi} - \frac{1}{4\pi} \log x_0\,.
  \end{aligned}
\end{equation}
Adding the two terms and for large $N$ one obtains
\begin{equation} 
  \begin{aligned} 
    I_{10} & =  \int_0^1 \mrd z \phi_0^2(z) + 
    \int_0^\infty \mrd z \left(\phi_0^2(z)-\frac{1}{4\pi z}\right)
    \\
    & \quad 
    +\frac{1}{4\pi}\int_0^1 \mrd x \left( S^2(x) - \frac{1}{x}\right)
    + \frac{1}{4\pi}\int_1^\infty \mrd x \left( S^2(x) - 1\right)
    \\ & \quad
    -\frac{1}{4 \pi} + \frac{1}{4\pi}\log\left(\frac{N^2}{4\pi} \right)
    + \order{1/N^2}\,.
  \end{aligned}
\end{equation}
Evaluating the integrals one has
\begin{equation} 
  I_{10} =  \frac{1}{2\pi} \log N + 0.048765633170141301742768467921 
  + \order{1/N^2}\,.
\end{equation}
For $\ell=1$ we trivially have
\begin{equation}
  I_{11} = \frac12 I_{00} = \frac12 - \frac{1}{2N^2}\,.
\end{equation}

For $n\ge2$ in the large $N$ limit the leading term of $I_{n0}$ 
is proportional to $N^{2n-2}$.  
The corresponding coefficient is given by (see \cite{Borwein})
\begin{equation}
  \lim_{N\to\infty} N^{2-2n} I_{n0} =
  \frac{1}{(2\pi)^{2n}} 
  \sum_{k_0,k_1=-\infty}^\infty\rule{0pt}{2.5ex}'\;
  \frac{1}{({\bf k}^2)^n}
  = \frac{4}{(2\pi)^{2n}}\zeta(n)\beta(n)\,,
\end{equation}
where $\zeta(n)$ and $\beta(n)$ are Riemann's zeta--function
and Dirichlet's beta--function, respectively.
In particular, up to $\order{N^{-2}}$ one has
\begin{equation}
  \frac{1}{V} I_{20} =  \frac{K}{24\pi^2} +\ldots 
  = 0.0038669465907372100307 + \ldots \,,
  \label{I20V}
\end{equation}
where $K=0.91596559417721901505$ is Catalan's constant
(since $\zeta(2)=\pi^2/6$ and $\beta(2)=K$).
One also has
\begin{equation}
  \frac{1}{V} I_{31} = \frac{1}{2V} I_{20} 
  = 0.0019334732953686050153 + \ldots \,.
  \label{I31V}
\end{equation}
Alternatively from the integral representation
\begin{equation}
  \frac{1}{V} I_{20}= \frac{1}{(4\pi)^2}\int_0^\infty \mrd x \, x
  \left[ S^2(x)-1\right] + \order{N^{-2}}\,,
  \label{I20VV}
\end{equation}
the evaluation of which agrees to all digits with \eqref{I20V}.

Further we have
\begin{equation}
  I_{21} = \frac12 I_{10}\,,
\end{equation}
and \cite{Shi97}
\begin{equation}
  I_{22;0} = \int_0^\infty \mrd z \, z \phi_0(z)\phi''_0(z) 
  = \frac12 - \frac{1}{2\pi}\,.
\end{equation}

Next
\begin{equation}
  \begin{aligned} 
    I_{32} & =  \frac12 \int_0^{x_0} \mrd z \, z^2 \phi_0(z) \phi''_0(z)
    + \frac{1}{8\pi} \int_{x_0}^\infty \mrd x\, x^2 S(x) S''(x)   \\
    & =
    \frac12 \int_0^1 \mrd z \, z^2 \phi_0(z) \phi''_0(z)
    + \frac12 \int_1^\infty \mrd z \, z^2 
    \left(\phi_0(z) \phi''_0(z)-\frac{3}{16\pi z^3}\right)   \\
    & \quad
    + \frac{1}{8\pi} \int_0^1 \mrd x \, x^2 
    \left(S(x) S''(x) - \frac{3}{4\pi x^3}\right) \\
    & \quad 
    + \frac{1}{8\pi} \int_1^\infty \mrd x \, x^2 S(x) S''(x) 
    + \frac{3}{32\pi}\log\left( \frac{N^2}{4\pi}\right) \\
    & = \frac{3}{16\pi}\log N + 0.0212534753416951596\,.
  \end{aligned}
\end{equation}
\begin{equation}
  I_{33;0} =  -\frac12 \int_0^\infty \mrd z \, z^2 \phi_0(z) \phi^{(3)}_0(z)
  = \frac12 - \frac{3}{4\pi}\,.
\end{equation}

\subsection{Some 2-loop momentum sums}

In this subsection we consider only the following 2-loop lattice
sums which appear in the computations \cite{Nie15a}:
\begin{align}
  W_{3a} &= -\sum_x G(x)^2 \nabla_0^2 G(x)\,, \label{W3a}
  \\
  W_{3c} & = \sum_x \nabla_0 G(x) \left[
    (\pz G(x))^2 - (\pz^* G(x))^2 \right]
   = - \frac16 \sum_x \left[\Box_0 G(x)\right]^3\,. \label{W3c}
\end{align}

It is sometimes useful to write
\begin{equation}
  W_{3a} = \widetilde{W}_{3a} - 2 G(0) \sum_x G(x) \nabla_0^2 G(x)\,,
  \label{W3ax1}
\end{equation}
where
\begin{equation}
  \widetilde{W}_{3a} = - \sum_x \left[G(x)-G(0)\right]^2 \nabla_0^2 G(x)\,.
  \label{W3ax2}
\end{equation}
Using $\nabla_0^2 = \Box_0 + \frac14 \Box_0^2$:
\begin{align} 
  \sum_x G(x) \nabla_0^2 G(x)&= -I_{21}+\frac14 I_{22}\,,
  \\
  \widetilde{W}_{3a} &= -\sum_x \left[G(x)-G(0)\right]^2 \Box_0 G(x)
  +\frac12 A_1\,,
\end{align}
where
\begin{equation}
  A_1=-\frac12\sum_x \left[G(x)-G(0)\right]^2 \Box_0^2 G(x)\,.
\end{equation}
For the symmetric case $\ell=1$ we have
\begin{equation}
  W_{3a}= I_{10}\left(\frac{1}{d}I_{10}-\frac12 I_{22}\right)
  -\frac{1}{dV}I_{20}+\frac12 A_1\,,\,\,\,\,\,\ell=1\,.
  \label{W3aelleq1}
\end{equation}

\subsubsection{Case $d=2$}

In this case we only consider the symmetric case where we can
use \eqref{W3aelleq1} and previous results to obtain
\begin{equation}
  W_{3a}=\frac{1}{8\pi^2}\ln^2 N
  +\frac{\ln N}{2\pi}\left( I_{10;0} - \frac14 + \frac{1}{4\pi}\right) 
  +W_{3a;0}+\order{1/N^2}\,,\,\,\,\,\,\ell=1\,,
  \label{W3a_d2}
\end{equation}
with
\begin{equation} 
  W_{3a;0}=\frac12 I_{10;0}\left(I_{10;0}-\frac12+\frac{1}{2\pi}\right)
  -\frac{K}{48\pi^2}
  +\frac 12a_1\,,
  \,\,\,\,\,\ell=1\,,
\end{equation}
where
\begin{equation}
  \begin{aligned}
    a_1 & = -\frac14 \int_{k,l} \frac{E_{k+l}-E_k-E_l}{E_k E_l E_{k+l}^2}
    \sum_\mu \widehat{(k+l)}^4_\mu \\
    & = 0.0461636292439177762(1)\,,
  \end{aligned}
\end{equation}
where $E_k=\hat{k}^2$. Putting in the numerical value for $I_{10;0}$
for $\ell=1$ we obtain
\begin{equation} 
  W_{3a;0}=0.0140266223093143915(1)\,,
  \,\,\,\,\,\ell=1\,.
\end{equation}

For the other double sum 
\begin{equation}
  W_{3c}=W_{3c;0}+\order{N^{-2}}\,,
\end{equation}
where $W_{3c;0}$ is the infinite volume sum which can 
be obtained for example by computing $W_{3c}$ for $\ell=1$:
\begin{align}
  -6W_{3c}(\ell=1)&=\frac12 \sum_x\left\{
    \left[\Box_0 G(x)\right]^3+\left[\Box_1 G(x)\right]^3\right\}
  \nonumber\\
  &=\frac12 \sum_x\left\{\Box_0 G(x)+\Box_1 G(x)\right\}
  \left\{\left[\Box_0 G(x)\right]^2-\Box_0 G(x)\Box_1 G(x)
    +\left[\Box_1 G(x)\right]^2\right\}
  \nonumber\\
  &=-\frac12\sum_x\left\{\delta(x)-\frac{1}{V}\right\}
  \left\{\left[\Box_0 G(x)\right]^2-\Box_0 G(x)\Box_1 G(x)
    +\left[\Box_1 G(x)\right]^2\right\}
  \nonumber\\
  &=-\frac18\left(1-\frac{1}{V}\right)^2+\frac{1}{2V}\sum_x
  \left\{\left[\Box_0 G(x)\right]^2-\Box_0 G(x)\Box_1 G(x)
    +\left[\Box_1 G(x)\right]^2\right\}
  \nonumber\\
  &=-\frac18+\order{1/V}\,.
\end{align}
So 
\begin{equation}
  W_{3c;0}=\frac{1}{48}\,,\,\,\,\,\,d=2\,.	
\end{equation}

\subsubsection{Case $d=3$}

We have determined the expansions by fitting the sums for a range
of values of $N$ for fixed $\ell$.
For the leading terms we have
\begin{align}
  W_{3a;0}&= 0.007958105980\,,\\
  W_{3c;0}&= 0.005702360998\,.
\end{align}
The large $N$ expansions are of the form:
\begin{align}
  W_{3a}[\ell=1]&= W_{3a;0} - 0.019432034237\frac{1}{N}
  + 0.013457106339\frac{1}{N^2} + \order{N^{-3}} \,,
  \label{W3a_l1_d3}
  \\
  W_{3a}[\ell=2]&= W_{3a;0} - 0.012367818370\frac{1}{N}
  + 0.012044486715\frac{1}{N^2} + \order{N^{-3}} \,, 
\end{align}
and
\begin{equation}
  W_{3c} = W_{3c;0} + \order{N^{-3}} \,.
\end{equation}

The coefficient $W_{3a;2}$ is related to 
$\psi,\beta_1,\overline{\beta}_1$ defined in \cite{Has93} through
\begin{equation}
  W_{3a;2} = \psi  + \frac13 \beta_1 \overline{\beta}_1\,.
  \label{W3a2_d3}
\end{equation}

\subsubsection{Case $d=4$}

Again we have obtained the expansions by fitting
the sums for a range of $N$ for fixed $\ell$.
For the infinite volume limit we obtained
\begin{align}
  W_{3a;0}&=0.001850792346021407306\,,
  \\
  W_{3c;0}&=0.00227599909081849438\,.
\end{align}
The expansion of $W_{3a}$ is given by 
\begin{align}
  W_{3a}[\ell=1]&= W_{3a;0} -0.004204473492568 \,N^{-2} 
  \nonumber\\
  & \qquad
  - (32\pi^2)^{-1} \,N^{-4} \ln N + 0.0077108092 \,N^{-4} + \ldots 
  \\
  W_{3a}[\ell=2]&= W_{3a;0} -0.0017695104622235 \,N^{-2} 
  \nonumber\\
  & \qquad
  + 0.0081775349546087202407 \,N^{-4} \ln N + 0.004168882181 \,N^{-4}
  + \ldots \\
  W_{3a}[\ell=3]&= W_{3a;0} +  0.00072483841\,N^{-2} 
  \nonumber\\
  & \qquad
  + 0.011346635653943 \,N^{-4} \ln N + 0.0131674 \,N^{-4}
  + \ldots \\
  W_{3a}[\ell=4]&= W_{3a;0} + 0.0032192873 \,N^{-2} 
  \nonumber\\
  & \qquad
  + 0.01292978432706041335 \,N^{-4} \ln N + 0.033262 \,N^{-4}
  + \ldots 
\end{align}
The coefficient of $N^{-4}\ln N$ can be obtained from the DR calculation.
This gives
\begin{equation}
  W_{3a;4x} =
  \frac{1}{48\pi^2}\left( \frac{1}{\ell} - 10 \ddot{g}(0;\ell)\right)\,.
  \label{W3a_4x}
\end{equation}
The fitting procedure reproduces this value.

For $W_{3c}$ we obtain:
\begin{align}
  W_{3c}[\ell=1]&= W_{3c;0} -0.01188332621 \,N^{-4} + \ldots \\
  W_{3c}[\ell=2]&= W_{3c;0} +0.01603789713 \,N^{-4} + \ldots \\
  W_{3c}[\ell=3]&= W_{3c;0} +0.02396645060 \,N^{-4} + \ldots \\
  W_{3c}[\ell=4]&= W_{3c;0} +0.027927570 \,N^{-4} + \ldots
\end{align}

The computation in \cite{Nie15a} required the following 
relations for consistency: 
\begin{align}
  W_{3a;2} & = \frac18 \left( 8\,I_{10;0}-1 \right) I_{10;2} \,, 
  \\
  W_{3a;4x} & = \frac{1}{48\pi^2} \left(10\,I_{11;4}
    +\frac{1}{\ell} \right) \,, 
  \\
  W_{3c;4} & =-\frac18\left(4\,I_{10;0} -1 \right)I_{11;4} \,.
\end{align}
These are satisfied by the numerical values above.

\section*{Acknowledgments}

We like to thank Christoph Weiermann for collaboration at an early stage
of the work. We would also like to thank Janos Balog, Peter Hasenfratz,
Heiri Leutwyler and Martin L\"uscher for useful discussions.

\begin{appendix}

\section{Derivation of formulae for $g(0)$ and 
$\ddot{g}(0)$ up to $\order{D-4}$ }
\label{AppA}

Using \eqref{g0dSA}, \eqref{Sddef} and \eqref{Phisub} we have
\begin{equation}
  g(0) L^{D-2} = \frac{1}{4\pi\mcVD} \int_0^\infty \mrd u\,
  \left[ \mcS_D \left(\frac{u}{\ell^2}\right)\right]_\mathrm{sub}
  -\frac{1}{4\pi\mcVD} - \frac{1}{2\pi(D-2)} \,.
\end{equation}
Expanding in $q$ one gets \eqref{g0xaA} and \eqref{g0xa1A}.

To calculate the second derivatives we use the relations
\begin{equation}
  \left.\frac{\partial^2}{\partial z^2}  S(v,z)\right|_{z=0} = 
  \sum_{n=-\infty}^\infty 
  \left( 4\pi^2v^2 n^2 - 2\pi v\right) \mre^{-\pi v n^2} 
  = -2\pi v \left( 2 v T(v) + 1\right) S(v) \,,
\end{equation}
\begin{equation}
  \left.\frac{\partial^2}{\partial x_\nu^2}  
    S\left(\frac{\ell_\nu^2}{u},\frac{x_\nu}{L_\nu}\right)\right|_{x_\nu=0} = 
  -\frac{2\pi l_\nu^2}{L^2 u} \, 
  \left[\frac{2\ell_\nu^2}{u}
    T\left(\frac{\ell_\nu^2}{u} \right) + 1\right]
  S\left(\frac{\ell_\nu^2}{u} \right) \,.
\end{equation}
From \eqref{gxdSA} one obtains then
\begin{equation}
  \begin{aligned}
    \partial_\nu^2 g(0) L^D & = - \frac12 \int_0^\infty \frac{\mrd u}{u^{D/2+1}}
    \left\{ 
      \left[ 
        \frac{2\ell_\nu^2}{u} T\left(\frac{\ell_\nu^2}{u}\right)
        + 1 \right]
      \mcS_D\left(\frac{\ell^2}{u}\right) -1 \right\}
    \\
    & = \int_0^\infty \mrd u
    \left\{ \frac{1}{\ell_\nu^2\mcVD}
      T\left(\frac{u}{\ell_\nu^2}\right)
      \mcS_D\left(\frac{u}{\ell^2}\right) 
      +\frac{1}{2u^{D/2+1}} \right\}
    \\
    & = \frac{1}{\ell_\nu^2\mcVD} \int_0^\infty \mrd u
    \left[ T\left(\frac{u}{\ell_\nu^2}\right)
      \mcS_D\left(\frac{u}{\ell^2}\right)
    \right]_\mathrm{sub} + \frac{1}{D} \,.
  \end{aligned}
\end{equation}
Expanding in $q$ one gets \eqref{ddg0xaA} and \eqref{ddg1xaA}.

\section{Evaluation of $\Psi$ for the massive case in DR}
\label{AppB}

Here we first considered the integral over the torus 
$\mathcal{T}=\prod_{\mu=0}^{d-1} S^1(L_\mu)[S^1(\Lhat)]^q$:
\begin{equation}
  \Wov(M) =
  -\int_{x\in\mathcal{T}} [G(x,M)]^2\partial_\nu^2G(x,M)\,,
  \,\,\,\,{\rm no\,\,sum\,\,over}\,\,\nu\,,
\end{equation}
with the massive free propagator $G(x,M)$.%

Using the Schwinger representation:
\begin{equation}
  \Wov(M) = \int_{x\in\mathcal{T}}
  \left[\prod_{j=1}^3\int_0^\infty\mrd t_j\right]\,\mcW\,,
\end{equation}
\begin{equation}
  \mcW =\mre^{-M^2(t_1+t_2+t_3)}\frac{1}{V_D}\sum_{p_1}     
  p_{1\nu}^2\exp\left\{ip_1x-p_1^2t_1\right\}
  \prod_{k=2}^3\left( \frac{1}{V_D}\sum_{p_k}
  \exp\left\{ip_kx-p_k^2t_k\right\} \right) \,.
  \label{mcW_Psi}
\end{equation}
We then break the $t_k$ integrations into parts ($t_0=L^2/(4\pi)$):
\begin{equation}
  \Wov=\Wov^{(1)}+2\Wov^{(2)}+\Wov^{(3)}+\Wov^{(4)}+2\Wov^{(5)}+\Wov^{(6)}\,,
\end{equation}
with
\begin{align}
  \Wov^{(1)}&=\int_{x\in\mathcal{T}}
  \int_{t_0}^\infty\mrd t_1\int_{t_0}^\infty\mrd t_2\int_{t_0}^\infty\mrd t_3\,
  \mcW\,,
  \\
  \Wov^{(2)}&=\int_{x\in\mathcal{T}}
  \int_{t_0}^\infty\mrd t_1\int_{t_0}^\infty\mrd t_2\int_0^{t_0}\mrd t_3\,
  \mcW\,,
  \\
  \Wov^{(3)}&=\int_{x\in\mathcal{T}}
  \int_0^{t_0}\mrd t_1\int_{t_0}^\infty\mrd t_2\int_{t_0}^\infty\mrd t_3\,
  \mcW\,,
  \\
  \Wov^{(4)}&=\int_{x\in\mathcal{T}}
  \int_{t_0}^\infty\mrd t_1\int_0^{t_0}\mrd t_2\int_0^{t_0}\mrd t_3\,
  \mcW\,,
  \\
  \Wov^{(5)}&=\int_{x\in\mathcal{T}}
  \int_0^{t_0}\mrd t_1\int_0^{t_0}\mrd t_2\int_{t_0}^\infty\mrd t_3\,
  \mcW\,,
  \\
  \Wov^{(6)}&=\int_{x\in\mathcal{T}}
  \int_0^{t_0}\mrd t_1\int_0^{t_0}\mrd t_2\int_0^{t_0}\mrd t_3\,
  \mcW\,,
\end{align}
and considered these separately. The advantage of this procedure
is that in each case the $x$ integration can be done first analytically.
The disadvantage over the coordinate space method is that
there are many terms to be considered and the integration
over the $t_i$ is often quite difficult.

The term $\Wov^{(1)}$ is the simplest:
\begin{align}
  \Wov^{(1)}(M)&=\frac{1}{V_D^2}
  \left[\prod_{j=1}^3\int_{t_0}^\infty\mrd t_j\right]
  \mre^{-M^2(t_1+t_2+t_3)}\sum_{p_1,p_2}p_{1\nu}^2
  \exp\left\{-p_1^2t_1-p_2^2t_2-(p_1+p_2)^2t_3\right\}
  \displaybreak[0]  \nonumber \\
  &=\frac{1}{V_D^2}\left(\frac{L^2}{4\pi}\right)^3\frac{4\pi^2}{L_\nu^2}
  \left[\prod_{j=1}^3\int_1^\infty\mrd t_j\right]
  \mre^{-(t_1+t_2+t_3)z^2/(4\pi)}
  \nonumber \\
  &\times\sum_{n_1,n_2}n_{1\nu}^2
  \exp\left\{-\sum_\mu\frac{\pi}{\ell_\mu^2}
    \left[n_{1\mu}^2t_1+n_{2\mu}^2t_2+(n_{1\mu}+n_{2\mu})^2t_3\right]\right\}
  \displaybreak[0]  \nonumber \\
  &=\frac{L^{4-2D}}{16\pi\mcV^2\ellhat^{2q}\ell_\nu^2}
  \sum_{m,n}m_\nu^2 R(z,m/\ell)R(z,n/\ell)R(z,(m+n)/\ell)\,,
\end{align}
where $z=ML$ and
\begin{align}
  R(z,v)&\equiv \frac{\exp\{-r(z,v)\}}{r(z,v)}\,, \label{Rzv}
  \\
  r(z,v)&\equiv \frac{z^2}{4\pi}+\pi v^2\,.
  \label{rzv}
\end{align}
Splitting off the terms in the double sum with $n=0$ and $n+m=0$
we can separate the singularities as $M\to0$ and 
the remaining sums can be computed accurately. 

Next we turn to $\Wov^{(2)}$; using the Poisson summation
formula (\ref{PF}) we obtain
\begin{align}
  \Wov^{(2)}(M)&=\frac{1}{V_D^2}\int_{x\in\mathcal{T}}
  \int_{t_0}^\infty\mrd t_1\int_{t_0}^\infty\mrd t_2\int_0^{t_0}\mrd t_3\,
  \mre^{-M^2(t_1+t_2+t_3)}
  \nonumber\\
  &\times (4\pi t_3)^{-D/2}
  \sum_n\exp\left(-\frac{\sum_\mu(x_\mu+n_\mu L_\mu)^2}{4t_3}\right)
  \nonumber\\
  &\times \sum_{p_1} p_{1\nu}^2\exp\left\{ip_1x-p_1^2t_1\right\}
  \sum_{p_2}\exp\left\{ip_2x-p_2^2t_2\right\}\,.
\end{align}

An important step now, which is also used for the remaining
$\Wov^{(a)}$,
is to use translation invariance to replace the
$x$ integration over $\mathcal{T}$ by an integration over
$\R^D$ which is readily performed: 
\begin{align}
  \Wov^{(2)}(M)&=\frac{1}{V_D^2}\int_{x\in\R^D}
  \int_{t_0}^\infty\mrd t_1\int_{t_0}^\infty\mrd t_2\int_0^{t_0}\mrd t_3\,
  \mre^{-M^2(t_1+t_2+t_3)}(4\pi t_3)^{-D/2}\exp\left(-\frac{x^2}{4t_3}\right)
  \displaybreak[0] \nonumber \\
  &\times \sum_{p_1}p_{1\nu}^2\exp\left\{ip_1x-p_1^2t_1\right\}
  \sum_{p_2}\exp\left\{ip_2x-p_2^2t_2\right\}
  \displaybreak[0] \nonumber \\
  &=\frac{1}{V_D^2}
  \int_{t_0}^\infty\mrd t_1\int_{t_0}^\infty\mrd t_2\int_0^{t_0}\mrd t_3\,
  \mre^{-M^2(t_1+t_2+t_3)}
  \nonumber \\
  &\times 
  \sum_{p_1,p_2}p_{1\nu}^2\exp\left\{-p_1^2t_1-p_2^2t_2-(p_1+p_2)^2t_3\right\}
  \displaybreak[0] \nonumber \\
  &=\frac{1}{V_D^2}\left(\frac{L^2}{4\pi}\right)^3\frac{4\pi^2}{L_\nu^2}
  \int_1^\infty\mrd t_1\int_1^\infty\mrd t_2\int_0^1\mrd t_3\,
  \mre^{-(t_1+t_2+t_3)z^2/(4\pi)}
  \nonumber \\
  &\times 
  \sum_{n_1,n_2}n_{1\nu}^2\exp\left\{-\sum_\mu\frac{\pi}{\ell_\mu^2}
    \left[n_{1\mu}^2t_1+n_{2\mu}^2t_2+(n_1+n_2)_\mu^2t_3\right]\right\}
  \nonumber \\
  &=\frac{L^{4-2D}}{16\pi\mcV^2\ellhat^{2q}\ell_\nu^2}
  \sum_{m,n}m_\nu^2 R(z,m/\ell)R(z,n/\ell)\overline{R}(z,(m+n)/\ell)\,,
\end{align}
where
\begin{equation}
  \overline{R}(z,v)\equiv \frac{1-\exp\{-r(z,v)\}}{r(z,v)}\,.
\end{equation}

The computation of all terms is too lengthy to present here.
The terms $\Wov^{(4)},\Wov^{(5)},\Wov^{(6)}$ have poles
at $D=4$ which are separated analytically. The remaining integrals
over the $t_i$ continued to $D=4$ are computable using e.g.
NAGLIB routines. 

Note that the singularities as $M\to 0$ in $\Wov(M)$ come from 
the terms $p=0$ in the propagators at finite volume. 
Our finite volume zero mass propagators exclude the zero mode.

\section{Some useful relations involving the lattice 
propagator}
\label{AppC}

Some other useful relations are:
\begin{equation}
  \begin{split}
    G(\hat{0}) & = I_{10} -  \frac12 I_{11}\,, 
    \\
    G(\hat{k}) & = I_{10} +  \frac{1}{2d_s}\left(I_{11}-I_{00}\right)\,,
    \\
    \pz G(0) & = -\frac12 I_{11}\,, 
    \\
    \pz^* G(0) & =   \frac12 I_{11}\,, 
    \\
    \partial_k G(0)& = -\frac{1}{2d_s}\left(I_{00}-I_{11}\right)\,,
    \\
    \partial_k^* G(0)& = \frac{1}{2d_s}\left(I_{00}-I_{11}\right)\,,
    \\
    \Box_0 G(0) & = -I_{11}\,, 
    \\
    \Box G(0) & = -I_{00}\,, 
    \\
    \pz^2 G(0) & =-I_{11}+\frac12 I_{12}
    \\
    \partial_k^2 G(0)& =
    \frac{1}{d_s}\left(-I_{00}+I_{11}-\frac12 I_{12}+\frac12 J_{10}\right)\,,
    \\
    \pz^*\partial_k G(0)& =-\frac{1}{4d_s}\left(I_{01}-I_{12}\right)\,,
    \\
    \pz\Box_0 G(0)& =\frac12 I_{12}\,, 
    \\
    \pz^*\partial_k^2 G(0)& =
    \frac{1}{2d_s}\left(-I_{01}+I_{12}-\frac12 I_{13}+\frac12 J_{11}\right)\,,
    \\
    \Box_0^2 G(0)& = I_{12}\,, 
    \\
    \Box\Box_0 G(0)& = I_{01}\,.
    \\
    \pz^*\pz^*\partial_k^2 G(0)& = \frac{1}{d_s}\left(I_{01}-I_{12}-\frac12
      I_{02}+I_{13}-\frac14 I_{14} -\frac12 J_{11} +\frac14 J_{12}\right)\,.
  \end{split}
  \label{G0_etc}
\end{equation}

\section{Modified Bessel function}
\label{AppD}

For integer order $n$ the modified Bessel function
can be calculated by
\begin{equation}
  I_n(x) = \sum_{k=0}^\infty  \frac{1}{k! (n+k)!} 
  \left( \frac{x}{2}\right)^{2k+n}\,.
  \label{In_ser}
\end{equation}

For $n\gg x$ one has
\begin{equation}
  I_n(x) \sim \frac{1}{n!} \left( \frac{x}{2}\right)^n 
  \sim \frac{1}{\sqrt{2\pi n}} \left( \frac{\mre x}{2n}\right)^n\,. 
  \label{Inx}
\end{equation}

For $\nu=\mathrm{fixed}$, $x\to\infty$:
\begin{equation}
  \mre^{-x}I_\nu(x) \sim  \frac{1}{\sqrt{2\pi x}}
  \sum_{k=0}^\infty (-1)^k \frac{a_k(\nu)}{x^k}
  \label{Ixn}
\end{equation}
where
\begin{equation} 
  a_k(\nu) =\frac{(4\nu^2-1)(4\nu^2-3^2)\ldots
    (4\nu^2-(2k-1)^2)}{k! \, 8^k}\,.
\end{equation}

For $n=0$ one has
\begin{equation}
  I_0(x) = \sum_{k=0}^\infty \frac{1}{(k!)^2} 
  \left( \frac{x}{2}\right)^{2k}\,,
  \label{I0_ser}
\end{equation}
and for $x\to\infty$
\begin{equation}
  \begin{aligned}
    \mre^{-x} I_0(x) & \sim  \frac{1}{\sqrt{2\pi x}}
    \sum_{k=0}^\infty  
    \left( \frac{(2k)!}{2^k (k!)^2}\right)^2 
    \frac{1}{k! (8x)^k}
    \\
    & = \frac{1}{\sqrt{2\pi x}}
    \left( 1 + \frac{1}{8x} + \frac{9}{128x^2}
      +\ldots
    \right) \,.
  \end{aligned}
  \label{Ix0}
\end{equation}
 
A uniform expansion for $I_\nu(\nu x)$ is 
$\nu\to\infty$
\begin{equation}
  \mre^{-\nu x} I_\nu(\nu x) \sim \frac{\mre^{\nu\rho}}{\sqrt{2\pi\nu}\,
    (1+x^2)^{1/4}} \sum_{k=0}^\infty \frac{U_k(p)}{\nu^k}\,,
  \label{Inxn}
\end{equation}
where
\begin{equation} 
  \begin{aligned}
    U_0(p) & =1\,,
    \\
    U_1(p) & =\frac{1}{24} (3p-5p^3)\,,
    \\
    U_2(p) & =\frac{1}{1152}(81p^2-462p^4+385p^6)\,,
  \end{aligned} 
\end{equation}
and
\begin{equation} 
  \begin{aligned}
    \rho & = \sqrt{1+x^2} - x +
    \ln\left(\frac{x}{1+\sqrt{1+x^2}}\right)\,, 
    \\
    p & = (1+x^2)^{-1/4}\,.
  \end{aligned} 
\end{equation}
Eq.~\eqref{Inxn} can also be
regarded as asymptotic expansion for 
$x\to\infty$, $\nu=\mathrm{fixed}$.

For large $x$ one has $\rho = -1/(2x) + \order{x^{-3}}$.
This shows that for $x_n \lesssim n^\alpha$ with $\alpha<2$
for $n\to\infty$ $\mre^{-x_n} I_n(x_n)$ decreases
exponentially fast.
Therefore
\begin{equation}
  N^{\alpha/2}\left[Q_N\left(x N^\alpha \right)
    -\phi_0\left(x N^\alpha \right) \right] = \order{\exp(-N^{2-\alpha}/x)} \,,
  \qquad 0\le \alpha<2 \,.
\end{equation}

Using \eqref{Ix0} one obtains in the large $N$ limit a simple scaling function,
\begin{equation}
  \lim_{N\to\infty} N^{\alpha/2} Q_N\left(x N^\alpha \right)
  = \frac{1}{\sqrt{4\pi x}}
  \,,
  \qquad 0<\alpha<2\,.
\end{equation}
For $\alpha=2$ this is not true any more, and according to
\eqref{SNlim} one obtains a non-trivial scaling function
\begin{equation}
  \lim_{N\to\infty} N Q_N\left(x N^2 \right) = S(4\pi x)\,.
\end{equation}

Summarizing one has three different scaling regimes
\begin{equation}
  \lim_{N\to\infty} N^{\alpha/2} Q_N\left(x N^\alpha \right) =
  \begin{cases}
    \phi_0(x) \,, & \alpha=0 \,, \\
    1/\sqrt{4\pi x} \, & 0 < \alpha < 2 \,, \\
    S(4\pi x) \,, & \alpha=2 \,.
  \end{cases}
\end{equation}

\end{appendix}



\begin{thebibliography}{10}

\bibitem{Nie15a}
  F.~Niedermayer and P.~Weisz,
  \emph{Matching effective chiral Lagrangians
   with dimensional and lattice regularizations},
  \emph{JHEP04} (2016) 110
   [arXiv:1601.00614]

\bibitem{Has90}
  P.~Hasenfratz and H.~Leutwyler,
  \emph{Goldstone boson related finite size effects in field 
       theory and critical phenomena with O($n$) symmetry},
  \emph{Nucl.\ Phys.\ B} {\bf 343} (1990) 241.

\bibitem{Bij13}
  J.~Bijnens, 
  \emph{Sunset integrals at finite volume},
  \emph{PoS\ LATTICE2013} (2014) 112
  [arXiv:1310.0350 [hep-lat]]. 

\bibitem{Bij14}
  J.~Bijnens, E.~Bostr\"{o}m and T.~A.~L\"{a}hde,
  \emph{Two-loop sunset integrals at finite volume},
  \emph{JHEP} {\bf 1401} (2014) 019
  [arXiv:1311.3531 [hep-lat]].

\bibitem{Lue91} 
  M.~L\"uscher, P.~Weisz and U.~Wolff, 
  \emph{A Numerical method to compute the running coupling in asymptotically 
    free theories},
  \emph{Nucl.\ Phys.\ B} {\bf 359} (1991) 221.

\bibitem{Has09}
  P.~Hasenfratz,
  \emph{The QCD rotator in the chiral limit},
  \emph{Nucl.\ Phys.\ B} {\bf 828} (2010) 201
  [arXiv:0909.3419 [hep-th]]

\bibitem{Lue86}
  M.~L\"{u}scher and P.~Weisz,
  \emph{Efficient numerical techniques for perturbative lattice gauge theory 
  computations},
  \emph{Nucl.\ Phys.\ B} {\bf 266} (1986) 309.

\bibitem{Lue95}
  M.~L\"{u}scher and P.~Weisz,
  \emph{Coordinate space methods for the evaluation of Feynman diagrams 
        in lattice field theories},
  \emph{Nucl.\ Phys.\ B} {\bf 445} (1995) 429.

\bibitem{Shi97} 	
  D-S.~Shin,
  \emph{Application of a coordinate space method for the evaluation 
        of lattice Feynman diagrams in two-dimensions},
  \emph{Nucl. \ Phys. } {\bf B525} (1998) 457.

\bibitem{Nec01} 	
  S.~Necco and R.~Sommer,
  \emph{The $N(f) = 0$ heavy quark potential from short to 
        intermediate distances},
  \emph{Nucl. \ Phys. } {\bf B622} (2002) 328.

\bibitem{Has93}
  P.~Hasenfratz and F.~Niedermayer,
  \emph{Finite size and temperature effects in the AF Heisenberg model},
  \emph{Z.\ Phys.\ B} {\bf 92} (1993) 91.

\bibitem{Borwein}
 J.~M.~Borwein and P.~B.~Borwein,
 \emph{Pi and the AGM: a study in analytic number theory 
       and computational complexity},
 \emph{New York: Wiley} 1987 p.~291.


\end{thebibliography}
\end{document}